\documentclass[journal, final, twocolumn]{IEEEtran} 
\usepackage{comment}
   \usepackage[pdftex]{graphicx} % Activated by CCD to be able to insert figures
   \graphicspath{{./figures/}} % Activated by CCD, to set the default figures directory
   \DeclareGraphicsExtensions{.pdf} % Activated by CCD for using PDFs only
\usepackage[cmex10]{amsmath} 
\usepackage{algpseudocode}
\usepackage{enumitem}
\usepackage{hyperref}

\usepackage{xcolor}  % COLOR. Remove when done. ?????

\ifCLASSOPTIONcompsoc
  \usepackage[caption=false,font=normalsize,labelfont=sf,textfont=sf]{subfig}
\else
  \usepackage[caption=false,font=footnotesize]{subfig}
\fi

\begin{document}
\title{Fast and Scalable Computation of the Forward and Inverse Discrete Periodic Radon Transform}

\author{Cesar~Carranza,~\IEEEmembership{Student Member,~IEEE,}
        Daniel~Llamocca,~\IEEEmembership{Senior Member,~IEEE,} \\
        and~Marios~Pattichis,~\IEEEmembership{Senior Member,~IEEE}% <-this % stops a space
\thanks{C. Carranza has been with the Department of Electrical and Computer Engineering
                   at the University of New Mexico.
He is currently with Secci\'on Electricidad y Electr\'onica, Pontificia Universidad Cat\'olica del Per\'u, Lima-32, Per\'u (e-mail: acarran@pucp.edu.pe)}
\thanks{M. Pattichis is with the Department
of Electrical and Computer Engineering, University of New Mexico, Albuquerque,
NM, 87131 USA (e-mail: pattichi@unm.edu)}% <-this % stops a space
\thanks{D. Llamocca is with the Electrical and Computer Engineering Department, Oakland University, Rochester, MI, 48309, USA (e-mail: llamocca@oakland.edu)}% <-this % stops a space

%\thanks{Manuscript received May 31, 2014; revised SOON!.}
}


\markboth{IEEE Transactions on Image Processing,~Vol.~XXX, No.~X, XXXX~XXXX}%
{Carranza \MakeLowercase{\textit{et al.}}: Fast and Scalable Architectures for the Forward and Inverse Discrete Periodic Radon Transform for Prime Sized Images}



\maketitle

\begin{abstract}
The Discrete Periodic Radon Transform (DPRT) has been extensively
   used in applications that involve image reconstructions from projections.
Beyond classic applications, the DPRT can also be used to compute
   fast convolutions that avoids the use of floating-point arithmetic
   associated with the use of the Fast Fourier Transform.   
Unfortunately, the use of the DPRT has been limited by the need to
   compute a large number of additions and the need for a large number
   of memory accesses.
   
This manuscript introduces a fast and scalable approach 
    for computing the forward and inverse DPRT
    that is based on the use of: 
     (i) a parallel array of fixed-point adder trees,
     (ii) circular shift registers to remove the need for accessing external memory components when selecting the input data for the adder trees,
     (iii) an image block-based approach to DPRT computation that can fit 
               the proposed architecture to available resources, and
     (iv) fast transpositions that are computed in one or a few clock cycles that
          do not depend on the size of the input image. 
As a result, for an $N\times N$ image ($N$ prime), 
    the proposed approach can compute up to $N^{2}$ additions per clock cycle.
Compared to previous approaches,    
          the scalable approach provides the fastest known implementations 
          for different amounts of computational resources.   
For example, for a $251\times 251$ image,  
             for approximately $25\%$ fewer flip-flops than required for a systolic implementation, 
             we have that the scalable DPRT is computed 
             36 times faster.             
For the fastest case, we introduce optimized architectures that can
            compute the DPRT and its inverse  
            in just $2N+\left\lceil \log_{2}N\right\rceil+1$ and  
            $2N+3\left\lceil \log_{2}N\right\rceil+B+2$ cycles respectively,
            where $B$ is the number of bits used to represent each input pixel.
On the other hand, the scalable DPRT approach requires
            more 1-bit additions than for the systolic implementation and
            provides a trade-off between speed and additional 1-bit additions.
All of the proposed DPRT architectures were implemented in VHDL and validated using an FPGA
  implementation.
\end{abstract}


% \fi
\IEEEpeerreviewmaketitle

\section{Introduction} \label{sec:intro}
%\IEEEPARstart{T}{his} demo file is intended to serve as a ``starter file''
%for IEEE journal papers produced under \LaTeX\ using
%IEEEtran.cls version 1.8 and later.

%Literaure review (Basic) - Do it historical: What they missed that you are addressing.
%
%Motivation (vs FFT): Primes have more sizes than power of two. DPRT is exact, FDPRT is faster, Scalable FDPRT!!!!.
%Why we care 

\IEEEPARstart{T}{he} Discrete Radon Transform (DRT) is an essential component of a wide range of applications in image processing \cite{jain1989,deans2007radon}.
   Applications of the DRT include the classic application of reconstructing objects from projections in computed tomography, radar imaging, and magnetic resonance imaging
      \cite{jain1989,deans2007radon}.
Furthermore, the DRT has also been applied in    
    image denoising \cite{Starck2002}, 
    image restoration \cite{Lun2004}, texture analysis \cite{Jafari2005}, 
    line detection in images \cite{Aggarwal2006}, and encryption \cite{Kingston2005}.
More recently, the DRT has been applied in
   erasure coding in wireless communications \cite{Normand2010},
   signal content delivery \cite{Parrein2012},
   and compressive sensing \cite{Ou2014}.

A popular method for computing the DRT involves the use of the 
  Fast Fourier Transform (FFT).
The basic approach is to sample the 2-D FFT 
   along different radial lines through the origin
   and then use the 1-D inverse FFT along each line to estimate the DRT.
This direct approach suffers from many artifacts that have been discussed in \cite{Starck2002}.
Assuming that the DRT is computed directly, Beylkin proposed an exact inversion algorithm
    in \cite{Beylkin1987}.
A significant improvement to this approach was proposed by 
    Kelley and Madisetti by eliminating interpolation calculations \cite{Kelley1993}.
A common  way to address this complexity 
    is to use Graphic Processing Unit (GPU) implementations as described in
    \cite{vlvcek2004}.
Unfortunately, this earlier work on the DRT requires the use of expensive floating point units
    for implementing the FFTs.
Floating point units require significantly larger amounts of hardware resources than
   fixed point implementations that will be discussed next. 

Fixed point implementations of the DRT can be based on the Discrete Periodic Radon Transform (DPRT). 
Grigoryan first introduced the forward DPRT algorithm for computing 
      the 2-D Discrete Fourier Transform as discussed in \cite{Grigoryan2010}.
In related work, Matus and Flusser presented a model for the DPRT and proposed 
    a sequential algorithm for computing the DPRT and its inverse for prime sized images 
    \cite{Matus1993}.
This research was extended by Hsung et al.
  for images of sizes that are powers of two \cite{Hsung1996}.

Similar to the continuous-space Radon Transform, 
   the DPRT satisfies discrete and periodic versions of the 
   the Fourier slice theorem and the convolution property.
Thus, the DPRT can lead to efficient, fixed-point arithmetic
   methods for computing circular and linear convolutions
   as discussed in \cite {Hsung1996}.
The discrete version of the Fourier slice theorem provides
   a method for computing 2-D Discrete Fourier Transforms
   based on the DPRT and a minimal number of 1-D FFTs
   (e.g., \cite{Gertner1988, Grigoryan2010}).

A summary of DPRT architectures based on the algorithm described by \cite{Matus1993}
  can be found in \cite{Ahmad2012}.
In \cite{Matus1993}, the DPRT of an image of size $N \times N$ ($N$ prime) 
  requires $(N+1)N(N-1)$ additions.
Based on the algorithm given in \cite{Matus1993}, a
  serial and power efficient architecture was proposed in \cite{Chandra2005}.
In \cite{Chandra2005},
  the authors used an address generator to generate the pixels to add.
The DPRT sums were computed using an accumulator adder
  that stores results from each projection using $N$ shift registers.
The serial architecture described in \cite{Chandra2005}
  required resources that grow linearly with the size of the image
  while requiring $N(N^{2}+2N+1)$ clock cycles to compute the full DPRT.

Also based on the algorithm given in \cite{Matus1993},     
  a systolic architecture implementation was proposed in \cite{Chandra2008}.
The architecture used a systolic array of $N(N+1)(\log_2 N)$ bits to 
  store the addresses of the values to add.
The pixels are added using using $(N+1)$ loop adder blocks.
The data I/O was handled by $N+1$ dual-port RAMs.
For this architecture, resource usage grows as $O(N^{2})$ 
  at a reduced running time of $N^{2}+N+1$ cycles for the full DPRT.
  
The motivation for the current manuscript is to investigate the development
    of DPRT algorithms that are both fast and scalable.
Here, we use the term {\it fast} to refer to the requirement that the 
    computation will provide the result in the minimum number of cycles.
Also, we use the term {\it scalable} to refer to the requirement that
     the approach will provide the fastest implementation based on
     the amounts of available resources.

This manuscript is focused on the case that the image is of size $N\times N$ and $N$ is prime.
   For prime $N$,
      the DPRT provides the most efficient implementations
      by requiring the minimal
      number of $N+1$ primal directions \cite{kingston2006projective}.
In contrast, there are $3N/2$ primal directions in the case that $N=2^p$ where $p$ is a positive
  integer \cite{pattichis2000novel}.
On the other hand, despite the additional directions, it is possible
  to compute the directional sums faster for $N=2^p$, as discussed
  in \cite{pattichis2000report,pattichis2001new}.
However, it is important to note that prime-numbered transforms
  have advantages in convolution applications.      
Here, just like for the Fast Fourier Transform (FFT),
  we can use zero-padding to extend the DPRT 
  for computing convolutions in the transform domain.   
Unfortunately, when using the FFT with $N=2^p$, zero-padding requires
  that we use FFTs with double the size of $N$.
In this case, it is easy to see that the use of prime-numbered DPRTs
  is better since there are typically many prime numbers between
  $2^p$ and $2^{p+1}$.   
For example,
  it can be shown that the $n$-th prime number
  is approximately $n\log(n)$ which 
  gives an approximate sequence of primes
  that are $n\log(n), (n+1)\log(n+1)$ 
  which is a lot more dense than
  what we can accomplish with powers of two $2^n, 2^{n+1}$
  \cite{hardy1979}.
As a numerical example, there are 168 primes that are less than 
  1000 as opposed to just 9 powers of 2.
Thus, instead of doubling the size of the transform, we can use
  a DPRT with only a slightly larger transform.

This manuscript introduces a fast and scalable approach 
    for computing the forward and inverse DPRT
    that is based on parallel shift and add operations.  
Preliminary results were presented in conference publications in
    \cite{carranza2014, carranza2014s}.
The conference paper implementations 
	 were focused on special cases of the full system discussed here, 
     required an external system to add the partial sums, assumed pre-existing
     hardware for transpositions, and worked with image strip-sizes that were limited to powers of two.
The current manuscript includes: 
   (i) a comprehensive presentation of the theory and algorithms,
   (ii) extensive validation that does not require external hardware for partial sums and transpositions,
   (iii) works with arbitrary image strip sizes, and 
   also includes (iv) the inverse DPRT.
In terms of the general theory presented in the current manuscript,
      the conference paper publications represented some special cases.
The contributions of the current manuscript over previously proposed approaches are
      summarized in the following paragraphs.

        Overall, a fundamental contribution of the manuscript is that it provides a
	    fast and scalable architecture that can be adapted to available resources.
         Our approach is designed to be fast in the sense that column sums are computed on every clock
            cycle.
         In the fastest implementation, a prime direction of the 
            DPRT is computed on every clock cycle.
         More generally, our approach is scalable, allowing us to handle larger images with
             limited computational resources.

        Furthermore, the manuscript provides a 
              Pareto-optimal DPRT and inverse DPRT based on running time and
 	            resources measured in terms of one-bit additions (or 1-bit full-adders) and flip-flops.
        Thus, the proposed approach is shown to be Pareto-optimal in terms of the required cycles and
           required resources.
        Here, Pareto-optimality refers to solutions that are optimal 
           in a multi-objective sense (e.g., see \cite{boyd2004convex}).
        Thus, in the current application, Pareto-optimality refers to the fact 
          that the scalable approach provides the fastest known implementations for the 
          given computational resources.
        As an example, in the fastest case,  for an $N\times N$ image ($N$ prime),
            we compute the DPRT in linear time ($2N+\left\lceil \log_{2}N\right\rceil+1$ clock cycles) 
            requiring resources that grow quadratically ($O(N^{2})$).
        In the most limited resources case, the running time is quadratic 
           ($\left\lceil N/2\right\rceil (N+9)+N+2$ clock cycles)   
            requiring resources that grow linearly  ($O(N)$).
        A Pareto-front of optimal solutions is given for resources that fall within these two extreme cases.
        All prior research in this area focused on the development of a single architecture.
        We also obtained similar results for the inverse DPRT, although 
          results for this case were not previously reported.
         
        In terms of speed, the manuscript describes the fastest 
             possible implementation of the DPRT and inverse DPRT.
       For the fastest cases, assuming sufficient resources for implementation,
           we introduce the fast DPRT (FDPRT) and the fast inverse DPRT (iFDPRT)
           that can compute the full transforms 
            in  $2N+\left\lceil \log_{2}N\right\rceil+1$ and  
            $2N+3\left\lceil \log_{2}N\right\rceil+B+2$ cycles respectively 
            ($B$ is the number of bits used to represent each input pixel). 
	
	To achieve the performance claims, we describe a
	    parallel and pipelined implementation that 
          provides an improvement over the sequential algorithm proposed by 
           \cite{Matus1993} and used in \cite{Chandra2005},\cite{Chandra2008}. 
        To summarize the performance claims, let the $N\times N$ input image 
             be sub-divided into strips of $H$ rows of pixels.
      Then, for $H=2, \dots, (N-1)/2$, our scalable approach computes $N\times H$ 
            additions in a single clock cycle.
     Furthermore, shift registers are used to make data available to the adders in every clock cycle.
     Then, additions and shifts are performed in parallel in the same clock cycle.

      In addition, we implement the use of fast transpositions.
     We propose two unique transpositions methods.
     First, we have a RAM-based architecture and associated algorithm that 
         provides a complete row or column of the input image in one clock cycle.
     Using this parallel RAM access architecture, transposition is avoided since 
         the image can be accessed by either rows or columns.
     Second, we use a register-based architecture that transpose the complete 
         register array in one clock cycle. This second approach avoids the use of RAMs.
    
     Finally, we provide a generic and parametrized family of architectures that is validated 
         with FPGA implementations. 
     Thus,  the proposed architectures are not tied to any particular hardware.
     They can be applied to any existing hardware (e.g., FPGA or VLSI)
         since they were developed in VHDL and are fully parametrized for any prime $N$. 

The rest of the manuscript is organized as follows.
The mathematical definitions for the DPRT and its inverse are given in section \ref{sec:background}.
The proposed approach is given in section \ref{sec:methods}.
 Section \ref{sec:implementation} describes the architecture implementation on a FPGA. Section \ref{sec:results} presents the results. Conclusions and future
work are given in section \ref{sec:conclusions}.

\section{Background}\label{sec:background}
The purpose of this section is to introduce the basic definitions
   associated with the DPRT and provide a very brief summary
   of previous implementations.
We introduce the notation in section \ref{sec:notation}.
We then produce the definitions of the DPRT and its inverse in 
    section \ref{sec:backDPRT}.
A summary of previous implementations is given in
    section \ref{sec:priorDPRT}.   

\subsection{Notation summary}\label{sec:notation}
We begin by introducing the notation.
We consider $N\times N$ images where $N$ is prime.
We let $Z_{N}$ denote the non-negative integers: $\left\{0, 1, 2, \dots, N-1\right\}$,
   and  $l^{2}(Z^{2}_{N})$ be the set of square-summable functions over $Z^{2}_{N}$.
Then, let $f \in l^{2}(Z^{2}_{N})$ be a 2-D discrete function that represents an 
      $N \times N$ image, where each pixel is a positive integer value represented with $B$ bits. 
Also, we use subscripts to represent rows.
For example, $f_k(j)$ denotes the
       vector that consists of the elements of $f$ where the value of $k$ is fixed.
Similarly, for $R(r,m,d)$, $R_{r,m}(d)$ denotes the
       vector that consists of the elements of $R$ with fixed values for $r, m$.   
Here, we note that we are always fixing all but the last index.             
We use $\left\langle\alpha\right\rangle_\beta$ to denote the modulo function.
In other words, $\left\langle\alpha\right\rangle_\beta$ 
    denotes the positive remainder when 
    we divide $\alpha$ by $\beta$ where $\alpha, \beta > 0$.

To establish the notation, we consider an example.
For an $251\times 251$ 8-bit image, we have $N=251$, $B=8$, and $f$ represents the image.
We then have that $f_1 (j)$ represents the first row in the image.
In 3-dimensions, $R_{1,2} (d)$ denotes the elements $R(r=1, m=2, d)$, where
   $d$ id allowed to vary.
For the modulo-notation, we have $\left\langle 255 \right\rangle_{251}=4$
    which represents the integral remainder when we divde 255 by $N=251$.  
We use $R(m,d)$ to denote the DPRT of $f$ 
    and $R'(r, m, d)$ to index the $r$-th partial sum
    associated with $R(m,d)$.
Here, $R'$ is used for explaining the computations associated with the scalable DPRT.

\subsection{Discrete Periodic Radon Transform and its Inverse}\label{sec:backDPRT}
We introduce the definition of the DPRT and its 
   inverse (iDPRT) based on \cite {Hsung1996}. 
Let $f$ be square-summable.
The DPRT of $f$ is also square summable and given by:
\begin{equation}  \label{eq:DPRT}
    R(m,d) = \left\{
    \begin{array}{ll}
	    \sum\limits^{N-1}_{i=0}f(i,\left\langle d+mi\right\rangle _{N}), & 0 \leq m < N, \\ % \limits forces the indices go on top or above of the sumation
	    & \\
	    \sum\limits^{N-1}_{j=0}f(d,j), & m = N, % \nolimits does the oposite
	  \end{array}
	  \right.
\end{equation}
where $d \in Z_{N}$ and $m \in Z_{N+1}$. 
In \eqref{eq:DPRT}, we observe that $m$ is used to index the prime directions
   as documented in \cite{carranza2014}.

The iDPRT recovers the input image as given by:
\begin{equation}  \label{eq:iDPRT}
    f\left(i,j\right) =\frac{1}{N}\left[\sum^{N-1}_{m=0}R\left(m,\left\langle j-mi\right\rangle_{N}\right)-S+R\left(N,i\right)\right]
\end{equation}
where:
\begin{equation}  \label{eq:fullsum}
    S =\sum^{N-1}_{j=0}\sum^{N-1}_{i=0}f(i,j).
\end{equation}
From \eqref{eq:fullsum}, it is clear that $S$
    represents the sum of all of the pixels.
Since each projection computes the sums over a single direction,
    we can sum up the results from any one of
    these directions to compute $S$ as given by:
\begin{equation}
   S = \sum^{N-1}_{d=0}R(m,d).
\end{equation}   

We note that the DPRT as given by \eqref{eq:DPRT} requires the computation of
  $N+1$ projections.
All of these projections are used in the computation of iDPRT 
  as given in \eqref{eq:iDPRT}.
In \eqref{eq:iDPRT}, the last projection computes $R(N, i)$ that is needed
  in the summation.
    
\subsection{DPRT implementations}\label{sec:priorDPRT}
DPRT implementations have focused on implementing the algorithm proposed in
    \cite{Matus1993}.
The basic algorithm is sequential that relies on computing the indices
    $i, j$ to access $f(i,j)$ that are needed for the additions in \eqref{eq:DPRT}.
For each prime direction, as shown in \eqref{eq:DPRT}, the basic implementation requires
   $N^2$ memory accesses and $N(N-1)$ additions. 
For computing all of the prime directions $(N+1)$, we thus
   have $(N+1)N^2$ memory accesses and $(N+1)N(N-1)$ additions.

Based on \cite{Matus1993}, hardware implementations have focused on
   computing memory indices, followed by the necessary additions
   \cite{Chandra2005}, \cite{Chandra2008}.
An advantage of the serial architecture given in \cite{Chandra2005}
   is that it requires hardware resources that grow linearly with $N$ (for and 
   $N\times N$ image).
Unfortunately, this serial architecture leads to slow computation
   since it computes the 
   DPRT in a cubic number of cycles ($N(N^{2}+2N+1)$ clock cycles).
A much faster, systolic array implementation was presented in \cite{Chandra2008}.
The systolic array implementation computes
   $N$ indices and $N$ additions per cycle.
Overall, the systolic array implementation requires
   hardware resources that grow quadratically with $N$
   while requiring $N^{2}+N+1$ clock cycles to compute the full DPRT.

The proposed architecture does not require memory indexing and computes
   the additions in parallel.
Furthermore, the new architecture is scalable, and thus allows
   us a consider a family of very efficient architectures that
   can also be implemented with limited resources.

\section{Methodology}\label{sec:methods}
This section presents a new fast algorithm and associated scalable architecture 
   that can be used to control the running time and hardware resources 
   required for the computation of the DPRT. 
Additionally, we extend the approach to the inverse DPRT (iDPRT).
At the end of the section, we provide an optimized architecture implementation
   that computes the DPRT and iDPRT in the least number of clock cycles.

% IMAGE SUBDIVISION IN STRIPS AND PARTIAL DPRT CALCULATION IDEA
\begin{figure}[!t]
\centering
\includegraphics[width=0.5\textwidth]{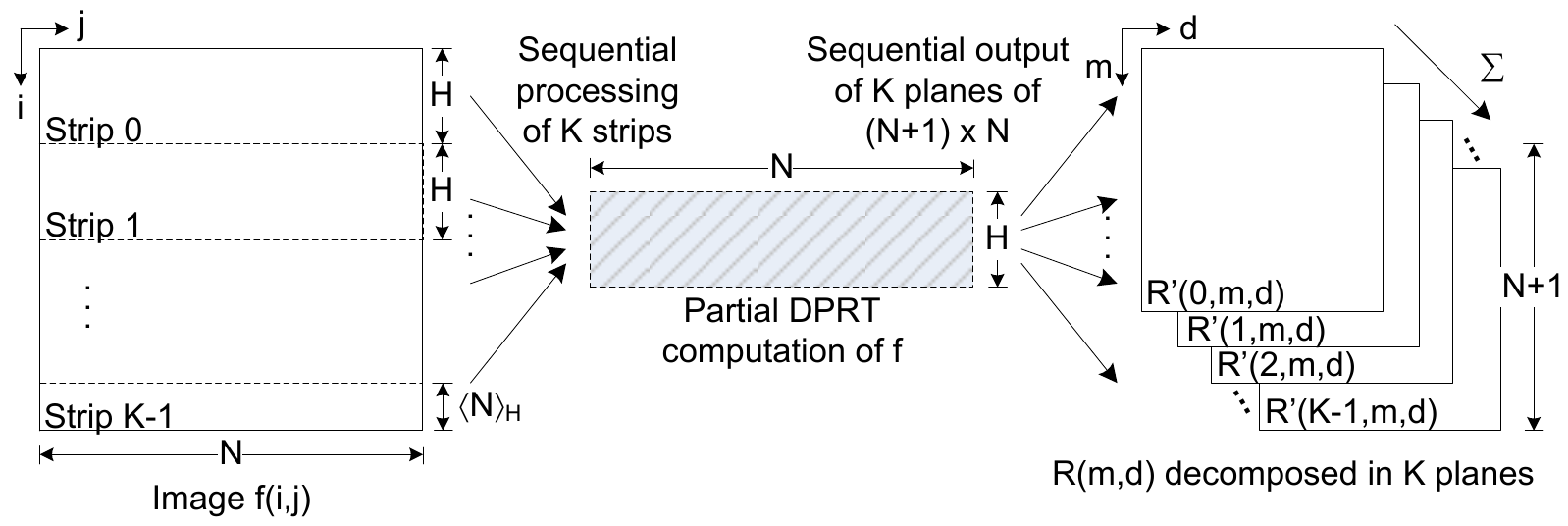} % Uses the whole column
\caption{Scalable DPRT concept.
	The input image is divided into $K$ strips.
	The DPRT is computed by accumulating the partial sums from each strip.
}
\label{fig:SFDPRTidea}
\end{figure}

\subsection{Partial DPRT}   
For the development of scalable architecture implementations, we introduce
    the concept of the partial DPRT.
The basic concept is demonstrated in Fig. \ref{fig:SFDPRTidea}
     and formally defined below.

The idea is to divide $f$ into strips that contain
     $H$ rows of pixels except for the last one that
     is composed of the remaining number of rows 
     needed to cover all of the $N$ rows
     (see Fig. \ref{fig:SFDPRTidea}).
Here, we note that the height of the last strip will be
   $\left\langle N\right\rangle _{H}\neq 0$ since $N$ is prime.
Now, if we let $K$ be the number of strips, we have that
    $K = \left\lceil N/H\right\rceil$.
In what follows, let $r$ denote the $r$-th strip.
We compute the DPRT over each strip using:
\begin{equation}
    R(m,d) = \left\{
    \begin{array}{ll}
	    \sum\limits^{K-1}_{r=0}\sum\limits^{L(r)-1}_{i=0}f(i+rH,\left\langle d+m(i+rH)\right\rangle _{N}), & \\
	    \hfill 0 \leq m < N & \\ % \limits forces the indices go on top or above of the sumation
	    & \\
	    \sum\limits^{K-1}_{r=0}\sum\limits^{L-1}_{j=0}f(d,j+rH), \hfill m = N &   % \nolimits does the oposite
	  \end{array}
	  \right.
\end{equation}
where
\begin{equation}
    L(r) = \left\{
    \begin{array}{ll}
	    H, & r < K-1\\
	    & \\
	    \left\langle N\right\rangle _{H} & r = K-1.  % \nolimits does the oposite
	  \end{array}
	  \right.
\end{equation}
We let $R'(r, m, d)$ denote the $r$-th partial DPRT defined by:
\begin{equation} \label{eq:partialDPRT}
    R' (r,m,d) = \left\{
    \begin{array}{ll}
	    \sum\limits^{L(r)-1}_{i=0}f(i+rH,\left\langle d+m(i+rH)\right\rangle _{N}), & \\
	    \hfill 0 \leq m < N & \\ % \limits forces the indices go on top or above of the sumation
	    & \\
	    \sum\limits^{L(r)-1}_{j=0}f(d,j+rH), \hfill m = N &   % \nolimits does the oposite
	  \end{array}
	  \right.
\end{equation}
where, $r=0, \ldots, K-1$ is the strip number. 
Therefore, the DPRT can be computed as a summation of partial DPRTs using:
\begin{equation}
    R (m,d) = \sum\limits^{K-1}_{r=0} R' (r,m,d).
\end{equation}

Similarly, we define the partial iDPRT of $R(m,d)$ using
\begin{equation} \label{eq:partialiDPRT}
    f'(r,i,j) = \sum\limits^{L(r)-1}_{m=0}R(m+rH,\left\langle j-i(m+rH)\right\rangle _{N})
\end{equation}
which allows us to compute the iDPRT of $R(m,d)$ 
  using a summation of partial iDPRTs:
\begin{equation} \label{eq:iSFDPRT}
    f(i,j) = \frac{1}{N}\left[\sum\limits^{K-1}_{r=0}f'(r,i,j)-S+R\left(N,i\right)\right].
\end{equation}
In what follows, we let let 
    $n = \left\lceil \log_{2}N\right\rceil$, 
    $h=\left\lceil \log_{2}H\right\rceil$, and
    $R'_{r,m}(d)$ be an $N$-th dimensional vector 
    representing the partial DPRT of strip $r$.

% SFDPRT TOP LEVEL HARDWARE
\begin{figure*}[!t]
\centering
\includegraphics[width=1.0\textwidth]{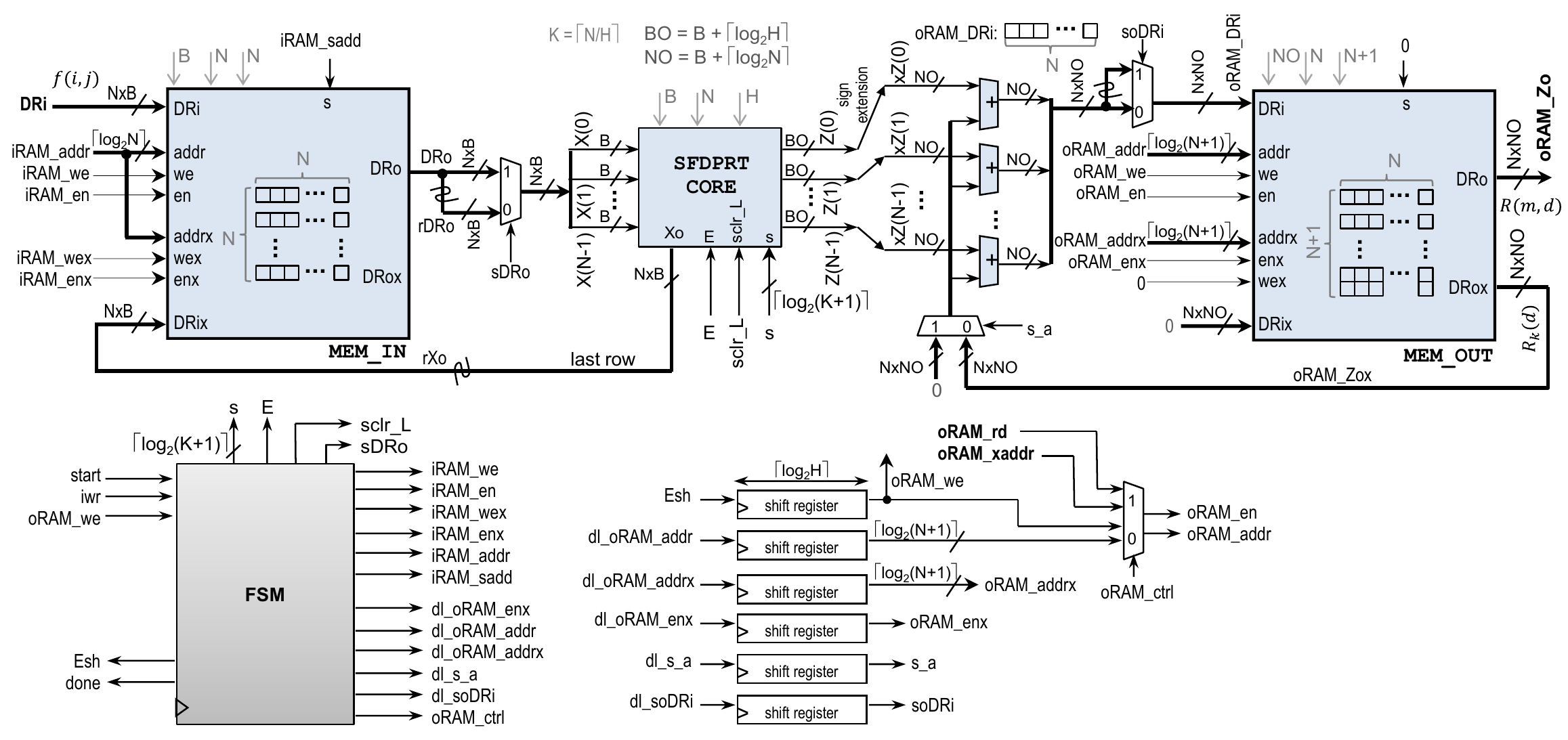} % Uses the whole column
\caption{
System for implementing the Scalable and Fast DPRT (SFDPRT).  
	  The ${\tt SFDPRT\_core}$ computes the partial sums.
	  ${\tt MEM\_IN}$ and ${\tt MEM\_OUT}$ are dual port input and output memories.
	  A Finite State Machine (${\tt FSM}$) is used for control. See text in Sec. \ref{subsec:scalable} for more details.
}
\label{fig:FPGASFDPRT}
\end{figure*}

\subsection{Scalable Fast Discrete Periodic Radon Transform (SFDPRT)}\label{subsec:scalable}
In this section, we develop the scalable DPRT hardware architecture
   by implementing the partial DPRT concepts presented in Fig. \ref{fig:SFDPRTidea}.
We present a top-level view of the hardware architecture for the scalable DPRT
   in Fig. \ref{fig:FPGASFDPRT}
   and the associated algorithm in Fig. \ref{alg:SFDPRTbasic}. 
We refer to Fig. \ref{fig:SFDPRTidea} for the basic concepts.     
The basic idea is to achieve scalability by controlling the number of rows used in each rectangular strip.
Thus, for the fastest performance,
    we choose the largest pareto-optimal strip size that can be implemented
    using available hardware resources. 
The final result is computed by combining the DPRTs as given in 
\eqref{eq:partialDPRT}.

We begin with an overview of the architecture as presented in Fig. \ref{fig:FPGASFDPRT}.
We have three basic hardware blocks:
   the input memory block (${\tt MEM\_IN}$),
   the partial DPRT computation block (${\tt SFDPRT\_core}$),
   and output/accumulator memory block (${\tt MEM\_OUT}$).  
The input image $f$ is loaded into the input buffer ${\tt MEM\_IN}$
   which can be implemented using a customized RAM 
   that supports access to each image row or column in a single clock cycle.
Partial DPRT computation is performed using the ${\tt SFDPRT\_core}$. 
We implement ${\tt SFDPRT\_core}$ 
   using an $H \times N$ register array with $B$ bits depth
   so as to be able to store the contents of a single strip.
Each row of the ${\tt SFDPRT\_core}$ register array is implemented
   using a Circular Left Shift (CLS) register 
   that can be used to align the image samples along each column.
Each column of this array has a $H$-operand fully pipelined adder tree 
   capable to add the complete column in one clock cycle. 
The output of the adder trees provide the output of the ${\tt SFDPRT\_core}$, 
   which represents the partial DPRT of $f$. 
This combination of shift registers and adders 
   allows the computation of $H \times N$ additions per clock cycle with a latency of $h$. 
At the end, the outputs of the ${\tt SFDPRT\_core}$ 
   are accumulated using ${\tt MEM\_OUT}$.
We summarize the required computational resources in
   section \ref{sec:results}.

% SFDPRT ALGORITHM
% htbp!
\begin{figure}[t]
\begin{algorithmic}[1]
\State \textbf{\textit{Load\_shifted\_image}} ($f$) in ${\tt MEM\_IN}$
                \label{step:SFDPRT1}
                
             using CLS registers of ${\tt SFDPRT\_core}$.     
\For {$r$ = $0$ to $K-1$} \label{step:SFDPRT2}
	\State \textbf{\textit{Load\_strip}}($r,$`row\_mode') into the ${\tt SFDPRT\_core}$ \label{step:SFDPRT6}
	\For {$k$ = $0$ to $N-1$} \label{step:SFDPRT7}
		\State Shift in parallel all the $H$ rows: \label{SFDPRT:shift}

 \ \ \ \ \ \ \ \ $CLS_{a}(H \cdot r+a)$, $a=0,\ldots,H-1$
		\State Compute in parallel $R^{'}_{r,k}(d)$ \label{SFDPRT:add}
		\State \textbf{\textit{Add\_partial\_result}}: $R_{k}(d)=R_{k}(d)+R^{'}_{r,k}(d)$ \label{SFDPRT:partial}
	
 \ \ \ \ \ \ \ \ in ${\tt MEM\_OUT}$
	\EndFor \label{step:SFDPRT8}
\EndFor \label{step:SFDPRT3}
\For {$r$ = $0$ to $K-1$} \label{step:SFDPRT4}
	\State $\textbf{\textit{Load\_strip}}$ ($r,$`column\_mode') into the ${\tt SFDPRT\_core}$
	\State Compute in parallel $R^{'}_{r,N}(d)$
	\State \textbf{\textit{Add\_partial\_result}}: $R_{N}(d)=R_{N}(d)+R^{'}_{r,N}(d)$
	
 \ \ \ \ in ${\tt MEM\_OUT}$
\EndFor \label{step:SFDPRT5}
\end{algorithmic}
\caption{\label{alg:SFDPRTbasic}
  Top level algorithm for computing the scalable and fast 
      DPRT (SFDPRT).
  Within each loop, all of the operations are pipelined.
  Then, each iteration takes a single cycle.
  For example, the Shift, pipelined Compute, and the Add operations of lines
      \ref{SFDPRT:shift}, \ref{SFDPRT:add}, and \ref{SFDPRT:partial} 
      are always computed within a single clock cycle.
   We refer to section \ref{sec:notation} for the notation.   
       }
\end{figure}

\begin{figure*}[!t]
\centering
\includegraphics[width=1.0\textwidth]{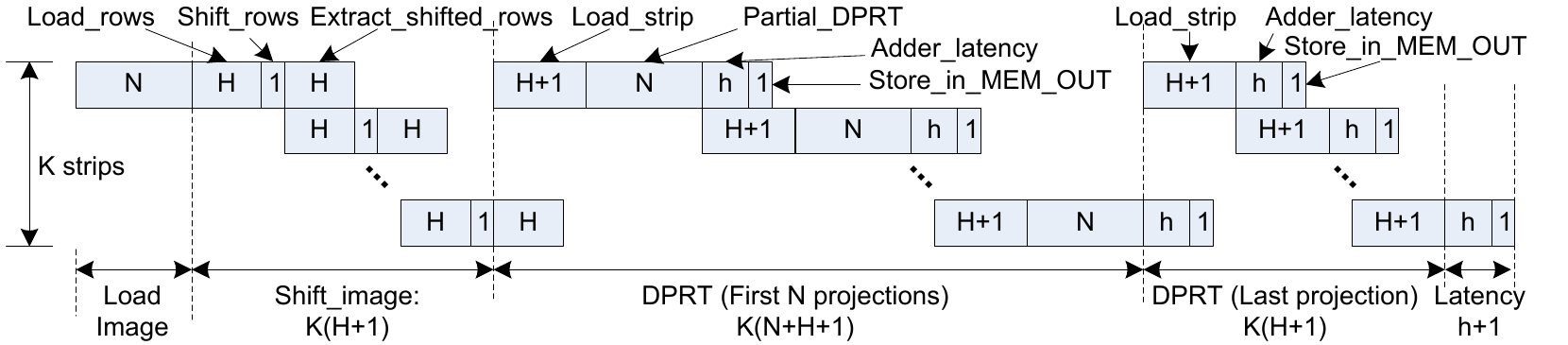} % Uses the whole column
\caption{Running time for scalable and fast DPRT (SFDPRT). 
	In this diagram, time increases to the right.
	The image is decomposed into $K$ strips.
	Then, the first strip appears in the top row and
	   the last strip appears in the last row of the diagram.
	Here, $H$ denotes the maximum number of image rows in each strip,
	      $K=\left\lceil N/H\right\rceil$ is the number of strips, and 
	      $h=\left\lceil \log_{2}H\right\rceil$ represents the addition latency.
}
\label{fig:SFDPRTtiming}
\end{figure*}

A fast algorithm for computing the DPRT is summarized in Fig. \ref{alg:SFDPRTbasic}.
We also present a detailed timing diagram for each of the steps in 
   Fig. \ref{fig:SFDPRTtiming}.
For the timing diagram, we note that time increases to the right.
Along the columns, we label each step and the required number of cycles.
Furthermore, computations that occur in parallel will appear along the same column.
To understand the timing for each computation,
  recall that $N$ denotes the number of image rows, 
  $K$ denotes the number of image strips where
  each strip contains a maximum of $H$ image rows.   
  
Furthermore, to explain the reduced timing requirements,
  we note the special characteristics of the pipeline structure.
First, we use dual port RAMs (${\tt MEM\_IN}$ and ${\tt MEM\_OUT}$)
  that allow us to load and extract one image row per cycle.
Thus, we start computing the first projection while we are still 
  shifting (also see the overlap between the second and third computing steps of Fig. \ref{fig:SFDPRTtiming}).
Second, we note that we are using fully pipelined adder trees
  which allow us to start the computation of the next projection
  without requiring the completion of the previous projection
  (see overlap in projection computations in Fig. \ref{fig:SFDPRTtiming}).

We next summarize the entire process depicted in Figs.
  \ref{alg:SFDPRTbasic} and \ref{fig:SFDPRTtiming}.
Initially, we load a shifted version of the image into ${\tt MEM\_IN}$.
The significance of this step is that the stored image allows 
    computation of the last projection in a single cycle without the need for transposition.
Here, we note that we can access rows and columns of ${\tt MEM\_IN}$ 
    in a single clock cycle.    
In terms of timing, the process of loading and shifting in the image requires
   $N+K(H+1)$ cycles.
     
Then, we compute the first $N$ projections by loading each one
  of the $K$ strips (outer loop) and then adding the partial results (inner loop).  
The partial DPRT for the strip $r$ is computed in the inner loop,
    (see lines \ref{step:SFDPRT7} - \ref{step:SFDPRT8}
     in Fig. \ref{alg:SFDPRTbasic}).
For computing the full DPRT, the partial DPRT outputs are accumulated
  in ${\tt MEM\_OUT}$. 
In terms of timing, each strip requires $N+H+1$ cycles as 
  detailed in Fig.  \ref{fig:SFDPRTtiming}.
Thus, it takes a total of $K(N+H+1)$ cycles for computing
  the first $N$ projections.

For the last projection, we note the requirement for special handling 
   (see lines \ref{step:SFDPRT4}-\ref{step:SFDPRT5} in Fig. \ref{alg:SFDPRTbasic}).        
This special treatment is due to the fact that unlike the first $N$ projections
that can be implemented effectively using shift and add operations of the rows,
the last projection requires shift and add operations of the columns.
For this last projection, we require $K(H+1)+h+1$ cycles
   which brings the total to  
   $K(N+3H+3)+N+h+1$ cycles for computing the full DPRT.
Furthermore, the DPRT is represented exactly
   by using $B+\lceil \log_2 N \rceil$ bits where
   $B$ represents the number of input bits.   

\subsection{Inverse Scalable Fast Discrete Periodic Radon Transform (iSFDPRT)}
\label{sec:iSFDPRT}
\begin{figure*}[!t]
\centering
\includegraphics[width=1.0\textwidth]{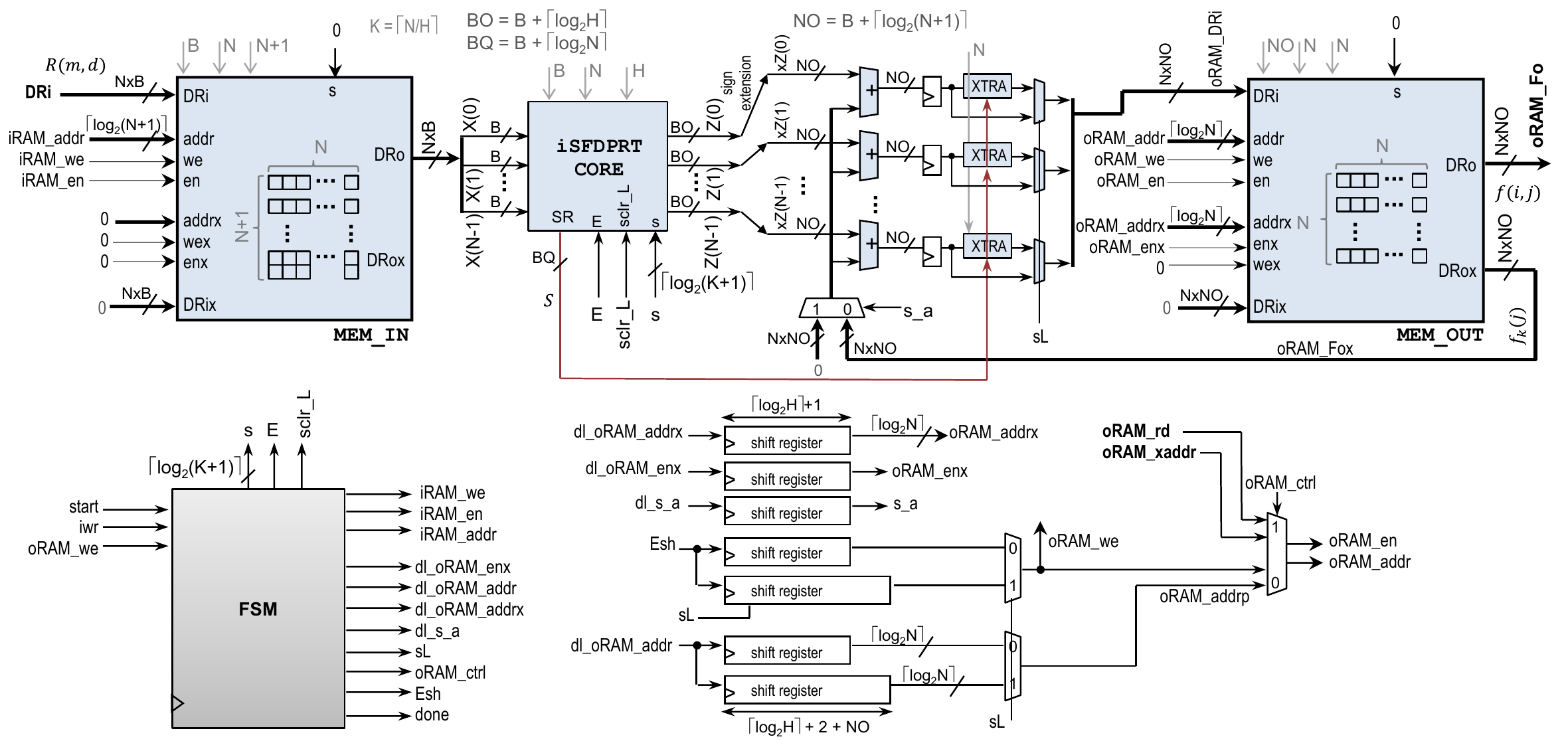} % Uses the whole column
\caption{System for implementing the inverse, scalable and fast DPRT (iSFDPRT).
	The system uses the ${\tt iSFDPRT\_core}$ core for computing partial sums. 
	The system uses dual port input and output memories, an accumulator array and a Finite State Machine for control.
	See text in Sec. \ref{sec:iSFDPRT} for more details.}
\label{fig:FPGAiSFDPRT}
\end{figure*}

The scalable architecture for the iDPRT is given in Fig \ref{fig:FPGAiSFDPRT},
   and the associated algorithm is given in Fig. \ref{alg:iSFDPRTbasic}.
 Here, we have three basic hardware blocks:
   (i) the input memory block (${\tt MEM\_IN}$),
   (ii) the partial inverse DPRT computation block (${\tt iSFDPRT\_core}$),
   and (iii) the output/accumulator memory block (${\tt MEM\_OUT}$).
 The functionality of this system is the same as the SFDPRT (see Sec. \ref{subsec:scalable}) with the 
   exception of the ${\tt XTRA}$ circuit that performs the normalization of the output.
Since there are many similarities between the DPRT and its inverse, we only
   focus on explaining the most significant differences.
The list of the most significant differences include:   
\begin{itemize}
  \item {\bf Input size:} The input is $R(m,d)$ with a size of $(N+1) \times N$ pixels.
  \item {\bf No transposition and optional use of ${\tt MEM\_IN}$:}
         A comparison between \eqref{eq:DPRT} and \eqref{eq:iDPRT} shows
         that second term of \eqref{eq:DPRT} is not needed for computing \eqref{eq:iDPRT}.
        Thus, the horizontal sums that required fast transposition are no longer needed.
        As a result, ${\tt MEM\_IN}$ is only needed to buffer/sync the incoming data.
        In specific implementations, ${\tt MEM\_IN}$ may be removed provided that
           the data can be input to the hardware in strips as described
           in Fig. \ref{alg:iSFDPRTbasic}.
  \item {\bf Circular right shifting replaces circular left shifts:}
        A comparison  between \eqref{eq:DPRT} and \eqref{eq:iDPRT} shows
           that the iDPRT index requires
              $\left\langle j-mi \right\rangle_N$
           as opposed to    
              $\left\langle d+mi \right\rangle_N$ 
           for the DPRT.   
        As a result, we have that the circular left shifts (CLS) of the DPRT
           become circular right shifts (CRS) for the iDPRT.
\end{itemize}

In terms of minor differences, we also note the special iDPRT terms
   of $R_{N}(d)$ and $S$ in \eqref{eq:iDPRT} that are missing from
   the DPRT.
These terms needed to added (for $R_{N}(d)$) and subtracted (for $S$)
   for each summation term.
Refer to Fig. \ref{alg:iSFDPRTbasic} for details.

\begin{figure}[t]
\begin{algorithmic}[1]
\For {$r$ = $0$ to $K-2$} \label{step:iSFDPRT2}
	\State Load strip $r$ into the ${\tt iSFDPRT\_core}$ \label{iSFDPRT:strip1}
	\If {$r = 0$}
	\State Compute $S$
  \EndIf
	\For {$k$ = $0$ to $N-1$} \label{step:iSFDPRT7}
		\State Shift in parallel all the $H$ rows: \label{step:iSFDPRT9}

 \ \ \ \ \ \ \ \ $CRS_{a}(H \cdot r+a)$,
 
 \ \ \ \ \ \ \ \ $a=0,\ldots,H-1$
		\State Compute in parallel $f^{'}_{r,k}(j)$ \label{alg:Z1}
		\State Add partial result $f_{k}(j)=f_{k}(j)+R^{'}_{r,k}(j)$
	
 \ \ \ \ \ \ \ \ in ${\tt MEM\_OUT}$
	\EndFor \label{step:iSFDPRT8}
\EndFor \label{step:iSFDPRT3}
	\State Load last strip into the ${\tt iSFDPRT\_core}$ \label{iSFDPRT:strip2}
	\For {$k$ = $0$ to $N-1$}
		\State Shift in parallel $\left\langle N\right\rangle _{H}$ rows: 

 \ \ \ \ \ \ \ \ $CRS_{a}(H \cdot r+a)$,
 
 \ \ \ \ \ \ \ \ $a=0,\ldots,\left\langle N\right\rangle _{H}-1$
		\State Compute in parallel $f^{'}_{r,k}(j)+R_{N}(d)$
		\State Add partial result: $f_{k}(j)=f_{k}(j)+f^{'}_{r,k}(j)+R_{N}(d)$
	  \State Subtract S: $f_{k}(j)=f_{k}(j)-S$
	  \State Normalize by N: $f_{k}(j)=f_{k}(j)/N$
	  
 \ \ \ \ \ \ \ \ and store in ${\tt MEM\_OUT}$
	\EndFor
\end{algorithmic}
\caption{\label{alg:iSFDPRTbasic}
{ Top level algorithm for computing the inverse Scalable Fast Discrete Periodic Radon
	 Transform $f(i,j)=\Re^{-1}(R(m,d))$.
  With the exception of the strip operations of lines \ref{iSFDPRT:strip1}
     and \ref{iSFDPRT:strip2}, all other operations are pipelined and executed
     in a single clock cycle.
  The strip operations require $H$ clock cycles where $H$ represents
      the number of rows in the strip.    	
  We refer to section \ref{sec:notation} for the notation.   
	}
}
\end{figure}

\begin{figure}[!t]
\centering
\includegraphics[width=0.5\textwidth]{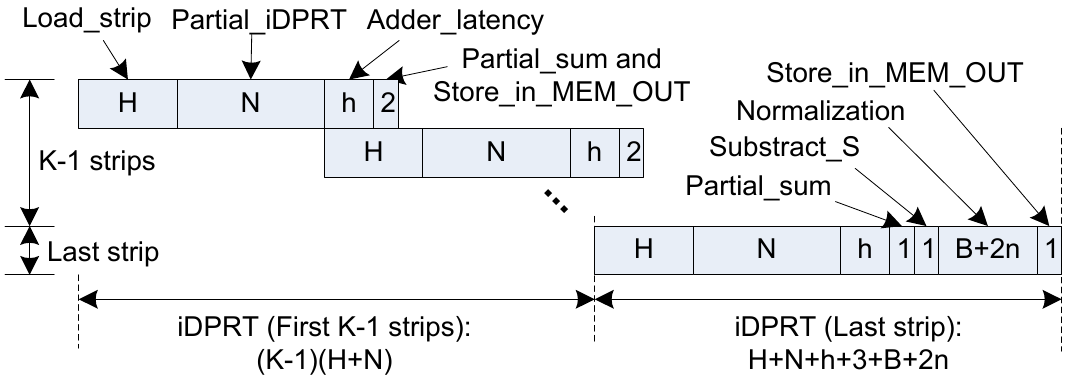} % Uses the whole column
\caption{Running time for computing the inverse, scalable, fast DPRT (iSFDPRT).
	Here, $H$ denotes the maximum number of projection rows for each strip, 
	      $K= \left\lceil N/H\right\rceil$ is the number of strips, 
	      $h=\left\lceil \log_{2}H\right\rceil$ is the addition latency, 
	      $n=\left\lceil \log_{2}N\right\rceil$, and $B+2n$ is the number of bits used to represent the results before normalization.
}
\label{fig:iSFDPRTtiming}
\end{figure}

We consider an optimized implementation that uses 
   pipelined dividers with a latency of as many clock cycles as 
   the number of bits needed to represent the dividend.
Then, the total running time is $K(N+H)+h+3+B+2n$ 
  as illustrated in Fig. \ref{fig:iSFDPRTtiming}. 
Resource requirements are given in 
  section \ref{sec:results}.

\subsection{Fast Discrete Periodic Radon Transform (FDPRT) and its inverse (iFDPRT)}
         \label{sec:FDPRT}
When there are sufficient resources to work with the entire image,
   there is no need to break the image into strips.
All the computations can be done in place
   without the need to compute partial sums that
   will later have to be accumulated.
In this case, we eliminate the use of the RAM
   and simply hold the input in the register array.               
For this case, we use the terms FDPRT and iFDPRT to describe
   the optimized implementations. 
For the FDPRT, the register array is also modified to implement 
   the fast transposition
   that is required for the last projection (transposition time=1 clock cycle).

\begin{figure*}
 \centering
 \includegraphics[width=0.98\textwidth]{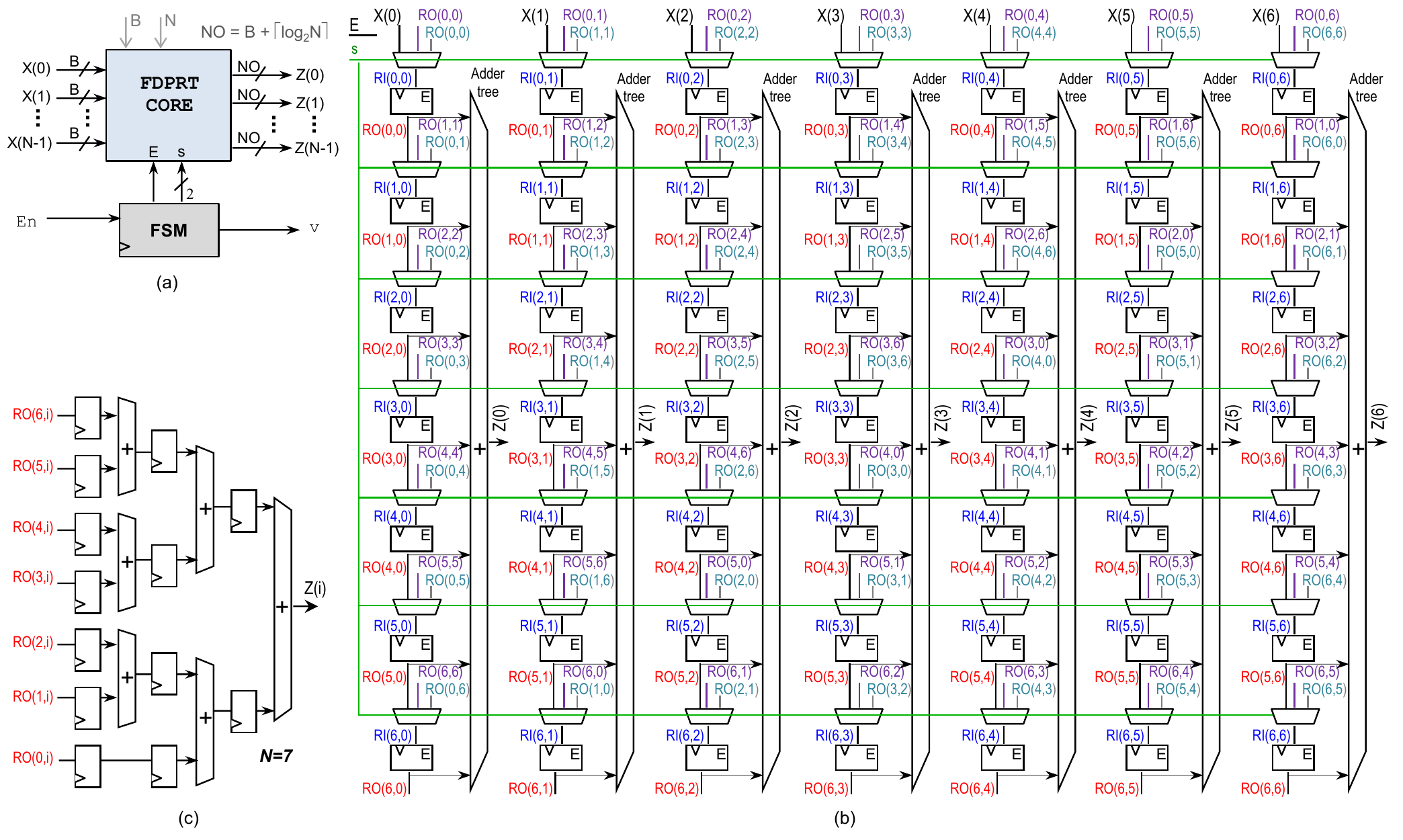}
 \caption{\label{fig:fpgasch}
			Fast DPRT (FDPRT) hardware.
        (a) FDPRT core and finite state machine (FSM).
        (b) Structure of the FDPRT core including: pipelined adder trees, registers,
              multiplexers (for shifting and fast transposition) for $N=7$.
        (c) Pipelined adder tree architecture for $N=7$.
   }
\end{figure*}

\begin{figure}[!h]
	\centering
	\includegraphics[width=0.5\textwidth]{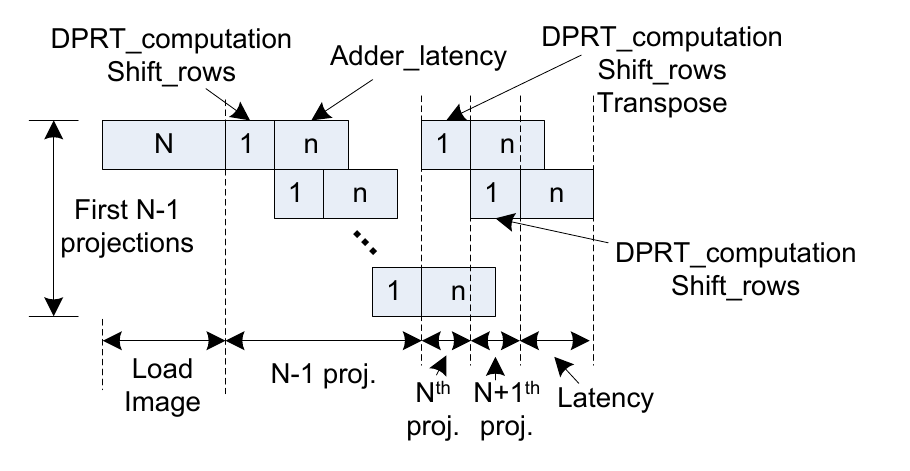} % Uses the whole column
	\caption{
		Running time for fast DPRT (FDPRT). 
		In this diagram, time increases to the right.
		The DPRT is computed in $N+1$ steps (projections). Each projection takes $1+h$ clock cycles.
		Here, $n=\left\lceil \log_{2}N\right\rceil$ represents the addition latency.
		Pipeline structure: Since we are using fully pipelined adder trees, the computation of 
		subsequent projections can be started after one clock of the previous projection.
	}
	\label{fig:FDPRTtiming}
\end{figure}

We present an example of the FDPRT hardware implementation for
   an $7\times 7$ image in Fig. \ref{fig:fpgasch}.   
We also present the associated timing diagram in Fig. \ref{fig:FDPRTtiming}.
Here, we note that time increases to the right.
As before, the different computational steps are depicted along the columns.
Cycles associated with parallel computations appear within the same column.

We next explain the process and derive the total running time in terms
  of the required number of cycles.
Initially, the image is loaded row-by-row at the top as shown in Fig. 
 \ref{fig:fpgasch}.
Thus, image loading requires $N$ cycles as depicted in the timing diagram
  of Fig. \ref{fig:FDPRTtiming}.
Shifting is performed in a single cycle along each row.
The shifted rows are then added along each column as shown in Fig. \ref{fig:fpgasch}.     
Due to the fully pipelined architecture, it only takes $N-1$ cycles
   to compute the first $N-1$ projections.
The last two projections only require two additional cycles.
Then, the final result is is only delayed by the latency
   associated with the last addition 
   ($n=\lceil\log_2 N \rceil$).    
Thus, overall, it only takes $2N+n+1$ cycles to compute the FDPRT.

For the iFDPRT, the architecture is basically reduced to the CRS registers 
   plus the adder trees.
For the iFDPRT, we also have    
an additional CLS(1), the subtraction of $S$ and the normalizing factor ($1/N$)
that is be embedded inside the adder trees. 
The iFDPRT algorithm is given in Fig. \ref{alg:iDPRT}. 
Overall, the iFDPRT requires $2N+3n+B+2$ cycles.
We summarize the required resources for both FDPRT and iFDPRT in section \ref{sec:results}.

\begin{figure}[h]
\begin{algorithmic}[1]
\State Load $R(m,d)$, \label{alg:loadflip}

 $0 \leq m,d \leq N-1$
\State Compute $S$ \label{alg:sumS}
\State Load $R_{N}(d)$
\State Compute in parallel $f_{0}(j)$

\For {$i$ = $1$ to $N-1$}
\State Shift in parallel the last $N-1$ rows

 \ \ \ \ $CRS_{k}(k), k=1,\ldots,N-1$
\State Compute in parallel $f_{i}(j)$ 
\EndFor
\end{algorithmic}
\caption{\label{alg:iDPRT}
{ Algorithm for computing the Inverse Fast Discrete Periodic Radon Transform $f(i,j) = \Re^{-1}(R(m,d))$}.
     We refer to section \ref{sec:notation} for the notation.
}
\end{figure}

\subsection{Pareto-optimal Realizations}\label{sec:Pareto}
For the development of scalable architectures,
   we want to restrict our attention to
   implementations that are optimal in the 
   multi-objective sense.
A similar approach was also considered in \cite{Llamocca2013}. \\~\\
Basically, the idea is to expect that
   architectures with more hardware resources
   will also provide better performance.
Here, we want to consider architectures
   that will give faster running times
   as we increase the hardware resources.       
   
The set of implementations that are optimal in 
   the multi-objective sense forms
   the Pareto front \cite{boyd2004convex}.
Formally, an implementation is considered to be sub-optimal 
  if we can find another (different) implementation that
  can run at the same time or faster 
  for the same or less hardware resources,
  excluding the case where both the running time and 
  computational resources are equal. 
The Pareto front is then defined 
   by the set of realizations that cannot be
   shown to be sub-optimal.

For deriving the Pareto-front, we fix the image size to $N$.
Then, we want to find the number of rows in each image
   strip (values of $H$) that generate Pareto-optimal architectures.
Now, since $N$ is prime, it cannot be divided by $H$ exactly.
The number of strips is given by
   $\lceil N/H \rceil$ which denotes the ceiling function
   applied to $N/H$.
To derive the Pareto-front, we require that larger values
   of $H$ will result in fewer strips to process.
In other words, we require that:
\begin{equation}
  \left\lceil \frac{N}{H}\right\rceil < \left\lceil \frac{N}{H-1}\right\rceil. 
  \label{eq:Pareto-front}
\end{equation}  
In this case, using $H$ rows in each strip will result in faster computations
   since we are processing fewer strips and we are also processing
   a larger number of rows per strip.
The Pareto front is then defined using:
\begin{equation}
  {\tt ParetoFront} = \left\{
     H\in S \quad\text{s.t.}\quad H \quad\text{satisfies eqn}\quad
     \eqref{eq:Pareto-front}
                      \right\} 
\end{equation}
where $S = \left\{2, 3, \dots, (N-1)/2 \right\}$ denotes
   the set of possible values for the number of rows.
To solve \eqref{eq:Pareto-front} and derive the
  ${\tt ParetoFront}$ set, we simply plug-in the different
  values of $H$ and check that \eqref{eq:Pareto-front}
  is satisfied.
Beyond the scalable approach, we note that an optimal architecture for $H=N$
   was covered in subsection \ref{sec:FDPRT}.
We will present the Pareto front in the Results section.

\section{Architecture implementation}\label{sec:implementation}
In this section, we will present the architecture implementations for
    the scalable and fast DPRTs and their inverses.
We provide a top-down description of the scalable architecture in section \ref{fpga:SFDPRT}.
Then, for the inverse DPRTs, we show the internal architecture that includes
    the circular shift registers and the adder trees.
Here, we note that the internal architectures for the forward DPRTs are closely related to the 
    architectures for the inverse DPRTs but simpler.    

\subsection{Scalable Fast Discrete Periodic Radon Transform (SFDPRT)}\label{fpga:SFDPRT}
\begin{figure}[!ht]
\centering
\includegraphics[width=0.5\textwidth]{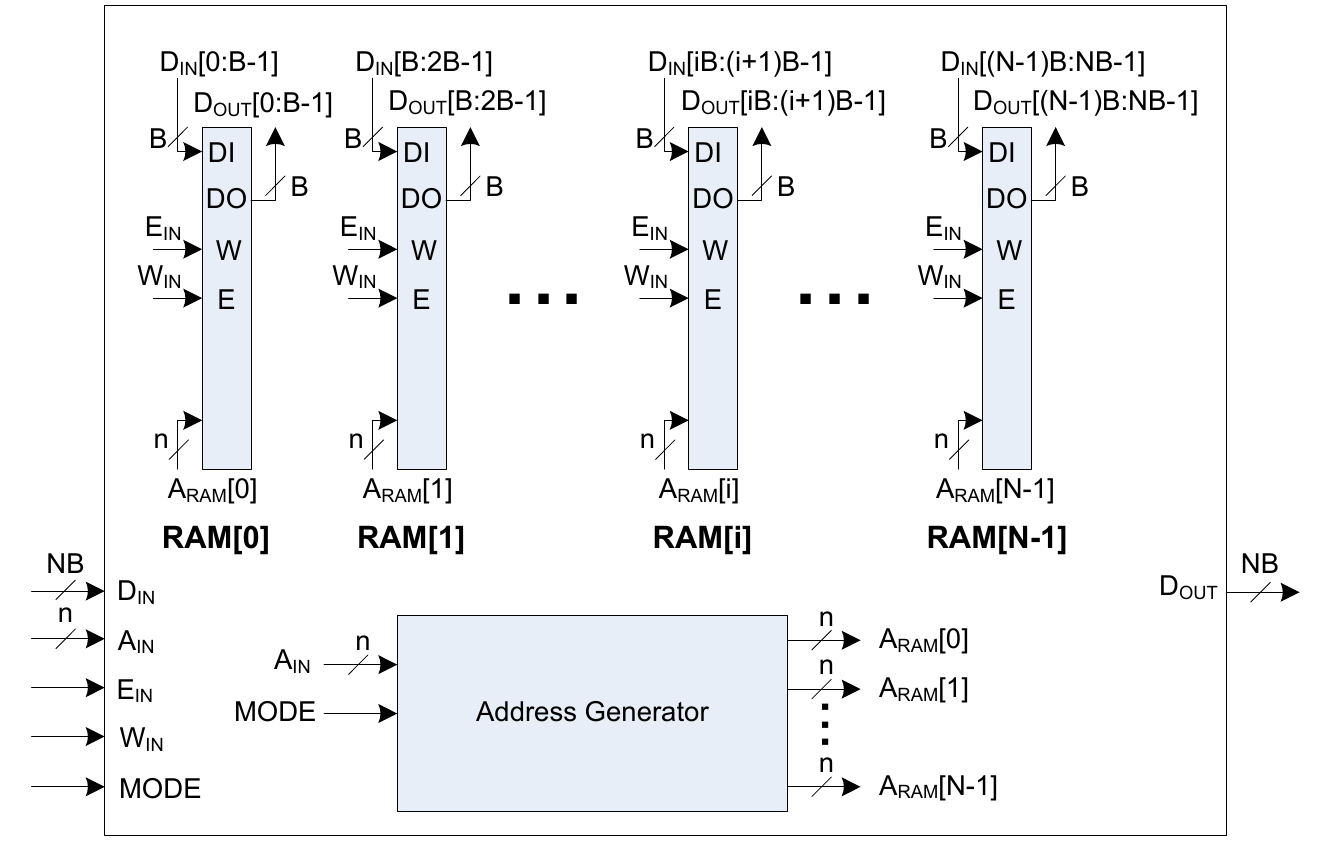} % Uses the whole column
\caption{Memory architecture for parallel read/write.
	     For the parallel load, refer to Fig. \ref{alg:LoadImage}.
	     The memory allows us to avoid transposition as described in Fig. \ref{alg:LoadBlock}.
	     The memory architecture refers to ${\tt MEM\_IN}$ and ${\tt MEM\_OUT}$
	        in Fig. \ref{fig:FPGASFDPRT}.
}
\label{fig:MEMIN}
\end{figure}

In this section, we will analyze and implement the different processes
  and components that were presented in the top-down
  diagram of Fig. \ref{fig:FPGASFDPRT}.
At the top level, we present  
  block diagrams for the the memory components (${\tt MEM\_IN}$ and ${\tt MEM\_OUT}$)
  in Fig. \ref{fig:MEMIN}.
We will conclude the section with FPGA implementations.

We begin with a brief description of the memory components.
Each {\tt RAM}-block is a standard Random Access Memory 
  with separate address, data read, and data write buses.
The \textit{MODE} signal is used to select between row and column access.
For row access, the addresses are set to the value stored in 
      ${\tt A_{RAM}\left[0\right]}$.
Column access is only supported for ${\tt MEM\_IN}$.
The addresses for column access are determined using:
 ${\tt A_{RAM}\left[i\right] = \left\langle A_{RAM}\left[0\right] + i\right\rangle_{N}, i = 1, \ldots ,N-1}$. 
 
 We summarize the main process of Fig. \ref{alg:SFDPRTbasic} in four steps:
\begin{enumerate}[itemindent=1cm,label=\bfseries Step \arabic*:]
	\item An $N \times N$ image is loaded row-wise in ${\tt MEM\_IN}$
	         as shown in line \ref{Load:theLoad} of Fig. \ref{alg:LoadImage}.
	\item Image strips are loaded into ${\tt SFDPRT\_core}$, shifted and written back
	         to ${\tt MEM\_IN}$  as described in Fig. \ref{alg:LoadImage}).
	          At the end of this step, the image is rearranged so that each diagonal corresponds 
	            to an image column. This allows us to get each row of the transposed image in one cycle.
	\item Image strips are loaded into the ${\tt SFDPRT\_core}$ and 
	           then left-shifted once as described in Fig. \ref{alg:LoadBlock}.
	           For the first $N$ projections, we accumulate the results from partial DPRTs computed
	                for each strip as described in Fig. \ref{alg:AddPartialResult}.
	           To compute the accumulated sums, we use an adder array.
	           Also, for pipelined operation, ${\tt MEM\_OUT}$ is implemented as a  dual port memory.
        \item   For the last projection, to avoid transposition, we access the input image in column mode.
                   The rest of the process is the same as for the previous $N$ projections. 
\end{enumerate}

\begin{figure}[!ht]
\begin{algorithmic}[1]
\State \textit{Load} image $f$ into ${\tt MEM\_IN}$ \label{Load:theLoad}	
\For {$z$ = $0$ to $K-1$}
	\For {$y$ = $0$ to $H-1$}
	    \State \textbf{\textit{Move}} row ($z*H+y$) of $f$ into ${\tt SFDPRT\_core}$
	    
 \ \ \ \ \ in reverse-order (flipped) \label{step:LoadImage3}
		\If {$z > 0$}
		    \State \textbf{\textit{Move}} the top row from ${\tt SFDPRT\_core}$ to
		    
 \ \ \ \ \ \ \ \ \ \ ${\tt MEM\_IN}$ in reverse-order at $((z-1)*H+y)$ \label{step:LoadImage6}
		\EndIf
	\EndFor
	\State \textbf{\textit {Shift}} in parallel all the $H$ rows 
	       into ${\tt SFDPRT\_core}$ \label{step:LoadImage7}
	
            registers: $CLS(z*H+a)$, $a=0,\ldots,H-1$.
\EndFor
\For {$y$ = $0$ to $H-1$}
	\State \textbf{\textit{Move}} the top row from ${\tt SFDPRT\_core}$ to

${\tt MEM\_IN}$ in reverse-order at $((K-1)*H+y)$ \label{step:LoadImage10}
\EndFor
\end{algorithmic}
\caption{\label{alg:LoadImage}
The implementation of \textbf{\textit{Load\_shifted\_image($f$)}} of Fig. \ref{alg:SFDPRTbasic}.
The process shifts the input image during the loading process in order to avoid
   the transposition associated with the last projection.
The shifting is performed using the circular left shift registers that are available
   in ${\tt SFDPRT\_core}$.  
}
\end{figure}

\begin{figure}[!h]
\begin{algorithmic}[1]
  \For {$z$ = $0$ to $H-1$}
        \If{$M==`row\_mode'$}
	\State \textbf{\textit{Move}} ${\tt MEM\_IN}$ row $(r \cdot H + z)$, mode $M$ \label{step:LoadBlock1}
	
	\ \ \ \ into ${\tt SFDPRT\_core}$. 
	
	\Else
	\State \textbf{\textit{Move}} ${\tt MEM\_IN}$ row $(r \cdot H + z)$, mode $M$ \label{step:LoadBlock2}
	
	\ \ \ \ into ${\tt SFDPRT\_core}$ in reverse-order (flipped).
	
	\EndIf
 \EndFor
 \State Shift in parallel all the $H$ rows: \label{step:SFDPRT9a}

 $CLS_{a}(H \cdot r+a)$, $a=0,\ldots,H-1$
\end{algorithmic}
\caption{\label{alg:LoadBlock}
Process for implementing $\textbf{\textit{Load\_strip(r, M)}}$ of Fig. \ref{alg:SFDPRTbasic}.
}
\end{figure}

\begin{figure}[t]
\begin{algorithmic}[1]
	\State \textbf{\textit{Read}} accumulated $R_{k}(d)$ from ${\tt MEM\_OUT}$
	\If {$k=N$}
		\State Flip $R^{'}_{k}(d)$
	\EndIf
	\State Add $R_{k}(d) = R_{k}(d) + R^{'}_{k}(d)$ 
	\State Store $R_{k}(d)$ in ${\tt MEM\_OUT}$    
\end{algorithmic}
\caption{\label{alg:AddPartialResult}
The implementation of  \textbf{\textit{Add\_partial\_result}} of Fig. \ref{alg:SFDPRTbasic}.
The process is pipelined where all the steps are executed in a single clock cycle.
}
\end{figure}

The Transform is computed using exact arithmetic using 
       $NO=B+\left\lceil \log_{2}N\right\rceil$ bits to represent the output
        where the input uses $B$-bits per pixel. 

\subsection{Inverse Discrete Periodic Transform Implementations}\label{fpga:iDPRT}
\subsubsection{iFDPRT}\label{fpga:iFDPRT}
\begin{figure*}[!t]
\centering
\includegraphics[width=1.0\textwidth]{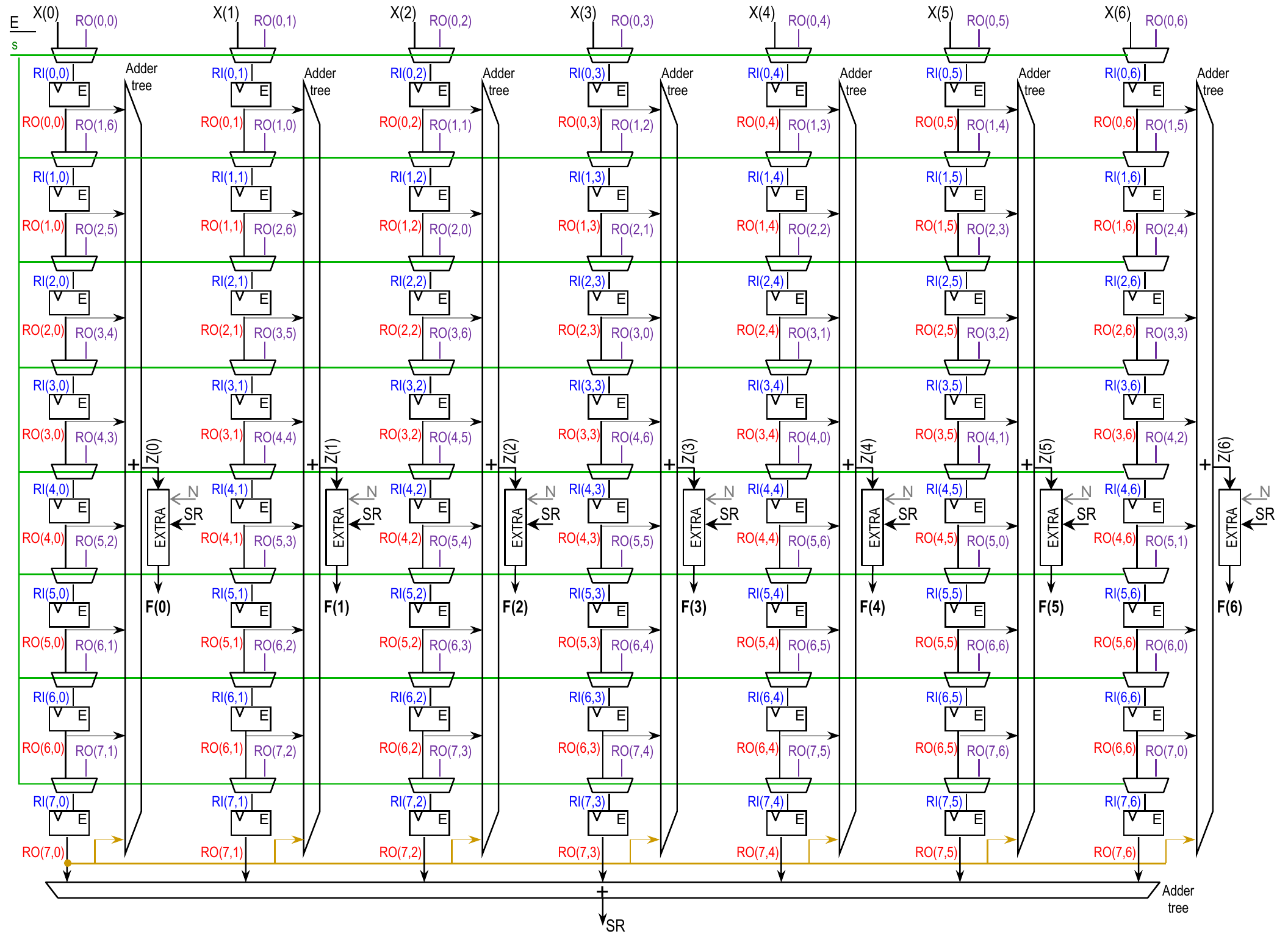} % Uses the whole column
\caption{The fast inverse DPRT (iFDPRT) hardware implementation.
The iFDPRT core shows the adder trees, register array, and 2-input MUXes.
Here, we note that the $Z(i)$ correspond to the summation term 
   in \eqref{eq:iDPRT} (also see Fig. \ref{alg:iDPRT}).
We note that the `extra circuit' is not needed for the forward DPRT.   
Also, for latency calculations, we note that the `extra circuit' has a latency of $1+BO$ cycles.    
}
\label{fig:FPGAiFDPRT}
\end{figure*}

We start by presenting the Inverse Fast Discrete Periodic Radon Transform (iFDPRT) core
    as shown in Fig. \ref{fig:FPGAiFDPRT}. 
The core generates an $N \times N$ output image based on an $(N+1) \times N$ input array. 
Fig. \ref{fig:FPGAiFDPRT} shows the array ($N=7$) where the shift is now to the right. 
Unlike the FDPRT core, we need to add $R(N,j)$ (an element of the last projection) for each computed direction $j$. We also need to subtract the sum $SR$ of a row and divide the result by $N$ 
    \eqref{eq:iDPRT}. We do not require transposition of the input array.
A total of $N$ directions are generated, where each direction is an $N$-element vector $F(i)$, $i=0, \ldots ,N-1$.

We next provide a brief overview of the different components.
We use 2-input MUXes to support loading and shifting as separate functions.
The vertical adder trees generate the $Z(i)$ signals. 
A new row  of $Z(0), Z(1), \ldots , Z(N-1)$ is generated for every cycle. 
The horizontal adder tree computes $SR$.
We recall that the $SR$ computation is the same for all rows as shown in \eqref{eq:fullsum}.
The latency of the horizontal adder tree is $\left\lceil \log_{2}N\right\rceil$ cycles. 
Note that $SR$ is ready when $Z(i)$ is ready, as the latency of the vertical adder trees is $\left\lceil \log_{2}(N+1)\right\rceil$. The $SR$ value is fed to the `extra units', where all $Z(i)$'s subtract $SR$ and then divide by $N$ (pipelined array divider \cite{Parhami2009} with a latency of $BO$ cycles).
The term $R(N,j)$ is included by loading the last input row on the last register row, where the shift is one to the left. Note that it is always the same element (the left-most one) that goes to all vertical adders. 

We also provide a summary of bitwidth requirements for \textit{perfect reconstruction}.
We begin by assuming that the Radon transform coefficients use $B'$-bits. 
The number of bits of the vertical adder tree outputs  
    $Z(i)$ are then set to $BO=B'+\left\lceil \log_{2}(N+1)\right\rceil$. 
The number of bits of $SR$ need to be $BQ=B'+\left\lceil \log_{2}N\right\rceil$. 
Assuming that the input image $f$ is $B$ bits, we only need $B$ bits to reconstruct it and
   the relationship between $B'$ and $B$ needs to be:
   $B' = B+\left\lceil \log_{2}N\right\rceil$ bits.
For the subtractor, note that
    $Z(i)= \sum^{N-1}_{m=0}R\left(m,\left\langle j-mi\right\rangle_{N}\right)+R\left(N,i\right)$
   and   then 
   $Z(i) \geq SR$ since $f(i,j)\geq 0$.
Thus, the result of $Z(i)-SR$ will always be positive requiring $BO$ bits. 
Thus, for perfect reconstruction, the result $F(i)$ needs to be represented using $BO$ bits.

\subsubsection{${\tt iSFDPRT\_core}$}\label{fpga:iSFDPRTcore}
\begin{figure*}[!t]
\centering
\includegraphics[width=0.9\textwidth]{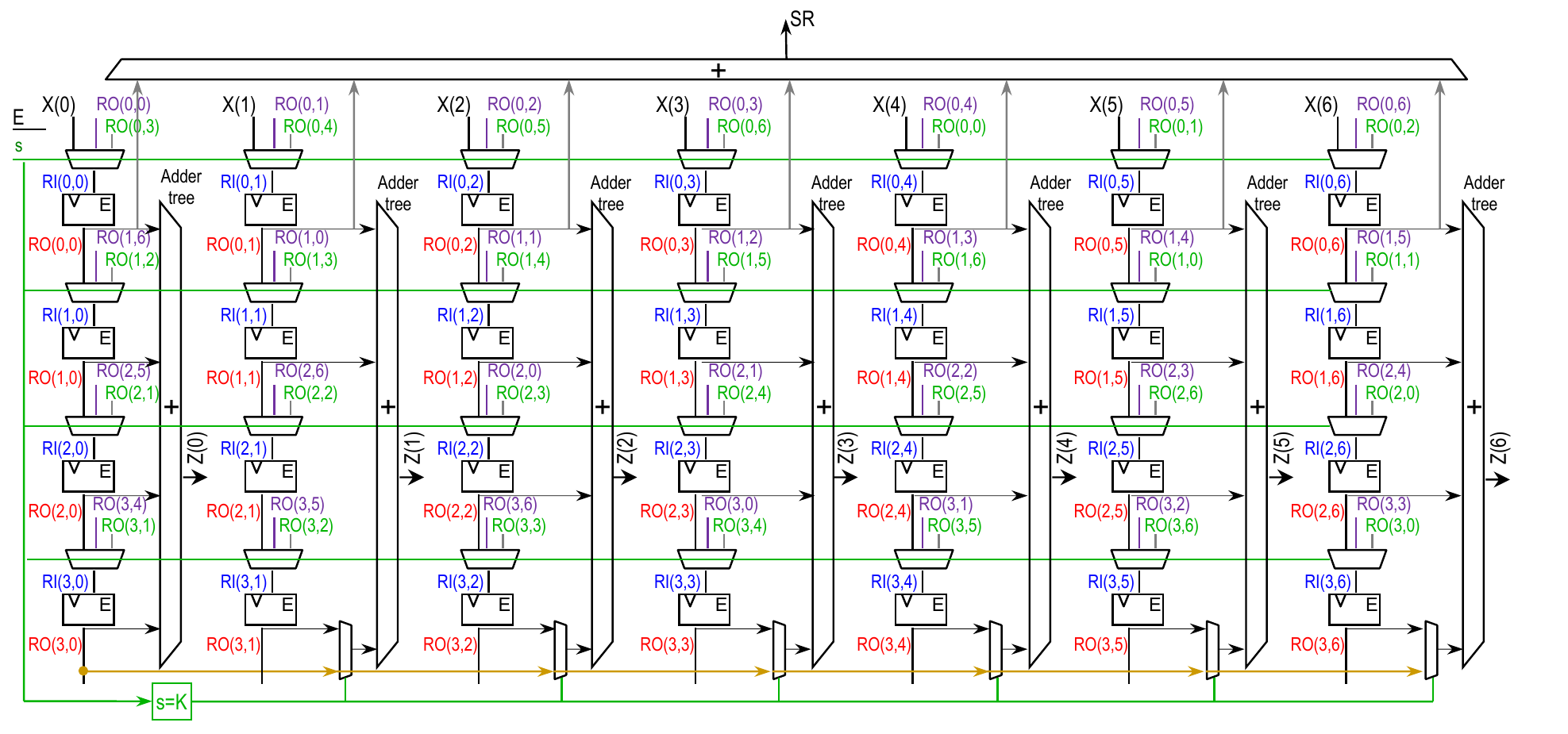} % Uses the whole column
\caption{The inverse scalable DPRT ${\tt iSFDPRT\_core}$ architecture for $N=7$, $H=4$.
% (b) Architecture for $N=7$, $H=3$. Note that here the last row that contains $R(N,j)$ is not in the last row of the last strip (which is loaded with zeros).
Here, we note that the $Z(i)$ correspond to the summation term in \eqref{eq:iSFDPRT}
    (also see Fig. \ref{alg:iSFDPRTbasic}).    
}
\label{fig:FPGAiSFDPRT_core_arch}
\end{figure*}

We next  summarize the core for the Inverse Scalable Fast Discrete Periodic Transform core 
      (${\tt iSFDPRT\_core}$).
Fig. \ref{fig:FPGAiSFDPRT_core_arch} shows an instance of this core for $N=7$ and $H=4$. 
The core only generates the partial sums $Z(i)$. 
We still need to accumulate the partial sums, subtract $SR$ from it and divide by $N$.

We next provide a summary of the required hardware.
For each strip, we need to be able to implement different amounts of right shifting.
This is implemented using $(K+1)$-input MUXes.
Since $N$ is always prime, we will have at-least one row of the register array 
   that will be unused during computations for the last strip.
The unused row is used to load the term $R(N,j)$.
The vertical MUXes located on the last valid row of the last strip ensure that the term 
   $R(N,j)$ is considered only when the last strip is being processed. 
Here, for the last row of the last strip, we require the shift to be one to the left.
Also, the remaining unused rows are fed with zeros.

\subsubsection{iSFDPRT}\label{fpga:iSFDPRT}
Recall that we presented the entire system in section \ref{sec:iSFDPRT}.
Beyond the ${\tt iSFDPRT\_core}$, we have the input and output memories, an array of adders and divisors, and ancillary logic. Here, we do not need to use the diagonal mode of the memories, flip the data, or rearrange the input memory. 
The basic process consists of loading each strip, processing it on the ${\tt iSFDPRT\_core}$, accumulating it to the previous result, and storing it in the output memory. For the last strip, we accumulate the result, but we also need to subtract $SR$ and divide by $N$.

\section{Results and Discussion}\label{sec:results}
\subsection{Results}
We next provide comprehensive results for both the scalable 
   and the fast DPRTs and their inverses.
We also compare the proposed approaches to previously
   published methods.   
      
\begin{figure}[!t]
\centering
\includegraphics[width=0.5\textwidth]{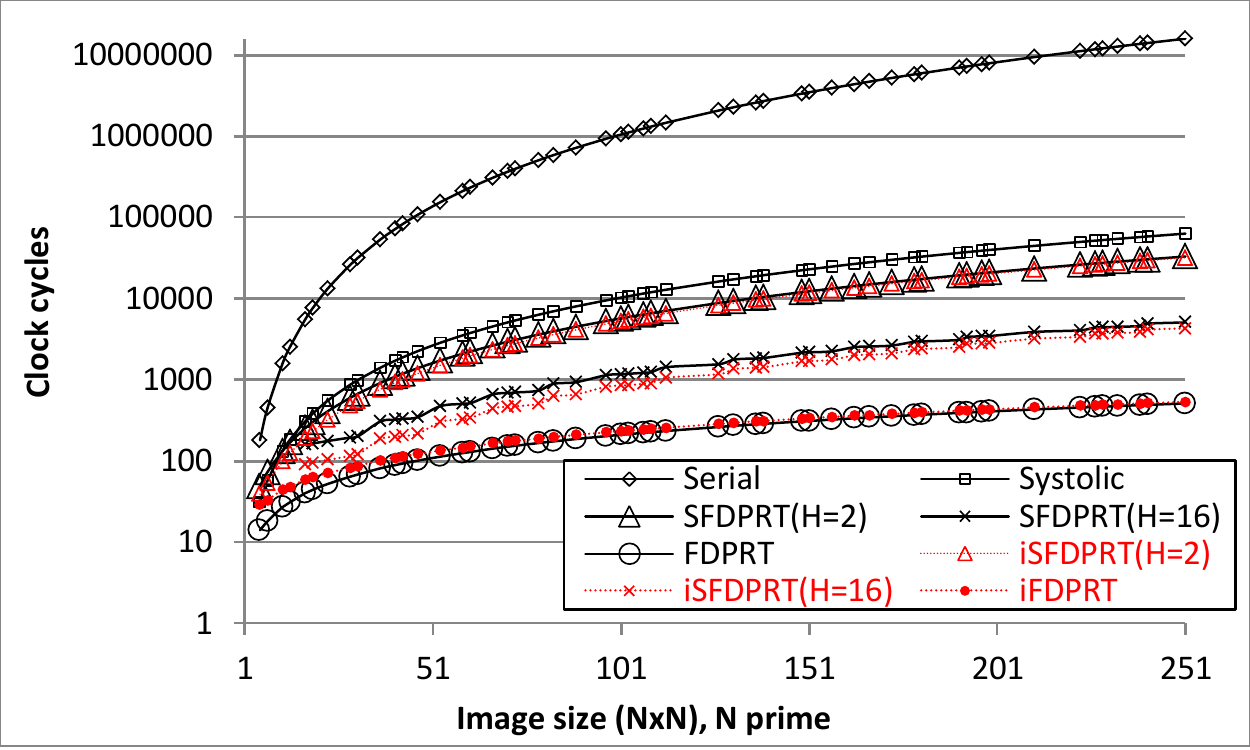} % Uses the whole column
\caption{Comparative running times for the proposed approach versus competitive methods.
We report running times in clock cycles for: 
  (i) the serial implementation of \cite{Chandra2005}, 
  (ii) the systolic \cite{Chandra2008},
   and (iii) the FPGA implementation of the SFDPRT for $H=2$ and $16$. 
The measured running times are in agreement with Tables \ref{table:timeforward}
     and \ref{table:timeinverse}.
}
\label{fig:FPGArunningTime}
\end{figure}

\begin{figure}[!h]
\centering
\includegraphics[width=0.5\textwidth]{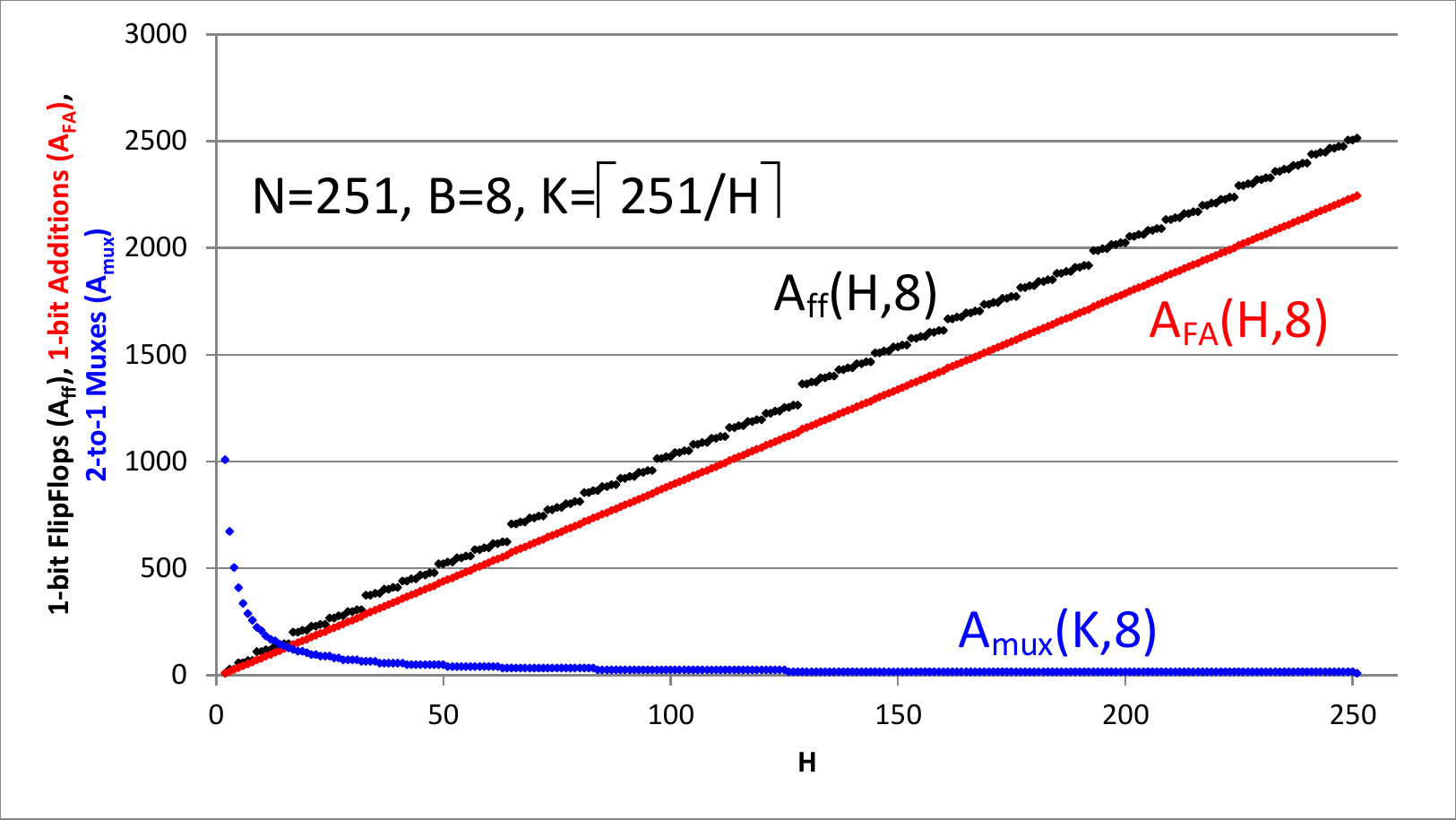} % Uses the whole column
\caption{
	 Resource functions: 
	 (i)   number of adder tree flip-flops ${\tt A_{ff}}(.)$,
	 (ii)  number of 1-bit additions ${\tt A_{fa}}(.)$, and
	 (iii) number 2-to-1 multiplexers ${\tt A_{mux}}(.)$ for $N=251$, $B$=8.
	Refer to Table \ref{table:resources} for definitions. 
}
\label{fig:JAffAfaAmux}
\end{figure}

\begin{figure}[!t]
\centering
\includegraphics[width=0.5\textwidth]{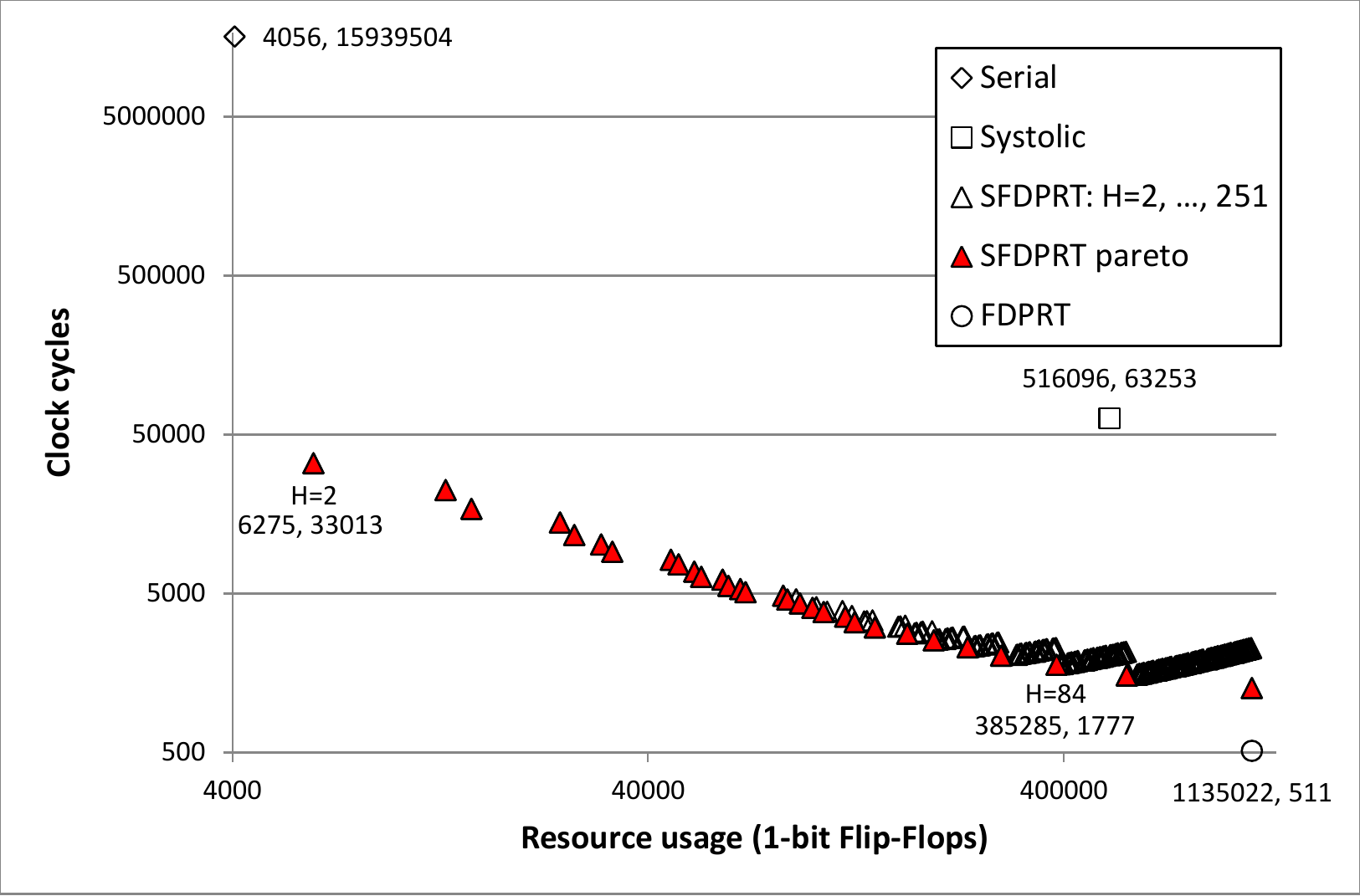} % Uses the whole column
\caption{
  Comparative plot for the different implementations based on
   the number of cycles and the number of flip-flops only.
  Refer to Fig. \ref{fig:JresourcesaddersTime} for a comparative
  plot for the different implementations based on the number of 
  cycles and the number of 1-bit additions.
Also, refer to Table \ref{table:Jresources_ram_mux_div} for a summary
  of RAM and multiplexer resources.
The plot shows the Pareto front for the proposed SFDPRT for $H=2, \ldots, 251$, for
  an image of size  $251 \times 251$.
The Pareto front is defined in terms of running time (in clock cycles) and 
  the number of flip-flops used. 
For comparison, we show the serial implementation from \cite{Chandra2005}, 
  and the systolic implementation \cite{Chandra2008}.
The fastest implementation is due to the FDPRT that is also shown.	
}
\label{fig:resourcesTime}
\end{figure}

\begin{figure}[!h]
	\centering
	\includegraphics[width=0.5\textwidth]{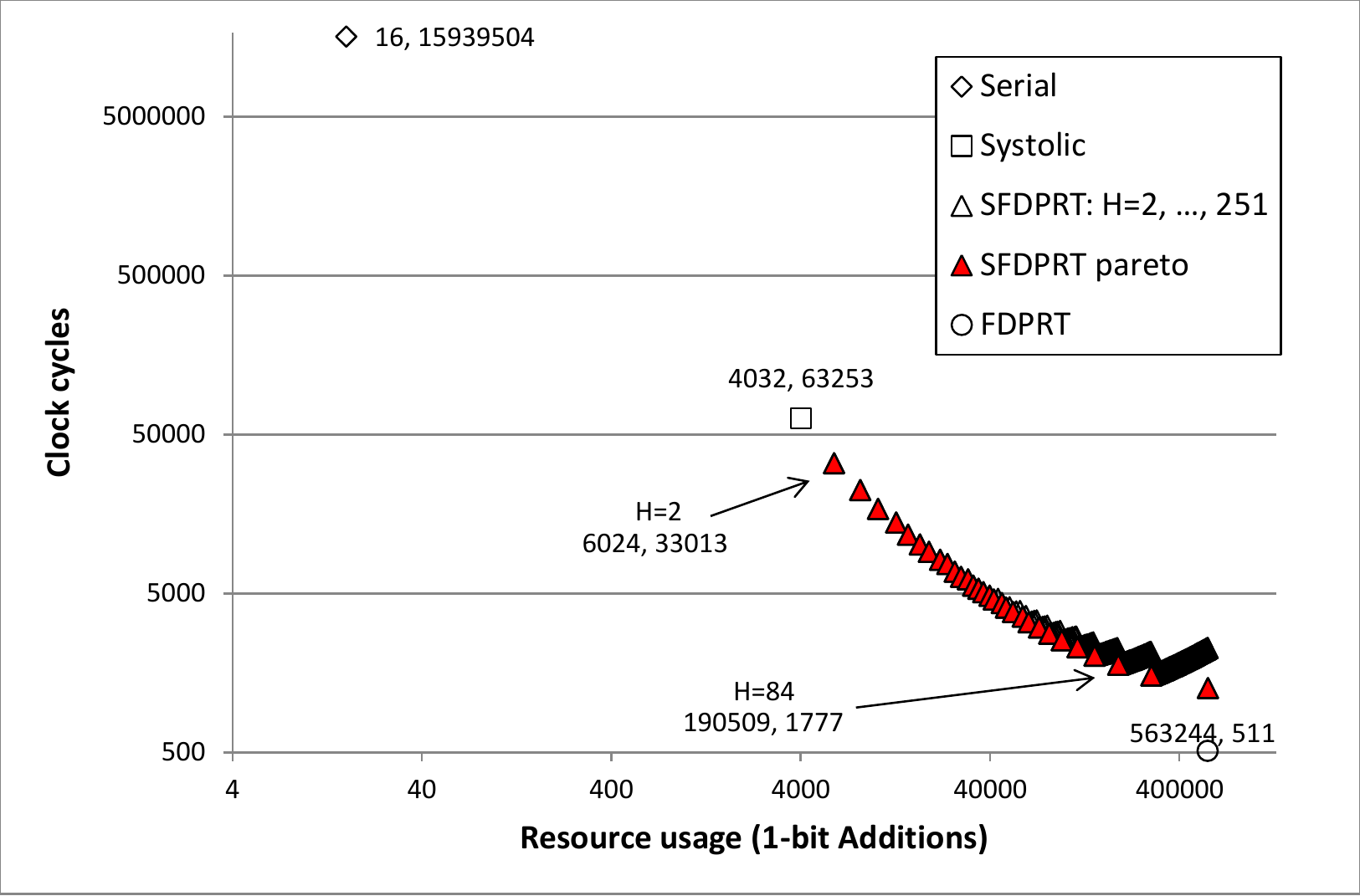} % Uses the whole column
	\caption{
		Comparative plot for the different implementations based on
		the number of cycles and the number of one-bit additions only
		   (or equivalent 1 bit full adders).
		Refer to Fig. \ref{fig:resourcesTime} for a similar comparison
		based on the number of flip-flops.
		Pareto front for the proposed SFDPRT for $H=2, \ldots, 251$, for
		an image of size  $251 \times 251$.
		The Pareto front is defined in terms of running time (in clock cycles) and the number of 
		1-bit additions. 
		For comparison, we show the serial implementation from \cite{Chandra2005}, 
		and the systolic implementation \cite{Chandra2008}.
		The fastest implementation is due to the FDPRT.
	}
	\label{fig:JresourcesaddersTime}
\end{figure}

We first present results as a function of image size. 
We summarize running times (in terms of the number of cycles) for the forward and inverse DPRT 
   in Tables \ref{table:timeforward}  and \ref{table:timeinverse} respectively.
For the forward DPRT, we compare against hardware implementations given by  
  the serial implementation in \cite{Chandra2005} and the systolic implementation in \cite{Chandra2008}.  
For the inverse DPRT, the computation times are similar. 
However, there are no exact values to compare against.
We also show comparative running times for $2<N<256$ for $B=8$ bits per pixel
  in Fig. \ref{fig:FPGArunningTime}, 

We provide a summary of the computational resources in Table \ref{table:resources}.
We also provide detailed
  resource functions ${\tt A_{ff}}$, ${\tt A_{FA}}$, and ${\tt A_{mux}}$ usage 
  for the 8-bit $251\times 251$ images in Fig. \ref{fig:JAffAfaAmux}.
In Fig. \ref{fig:JAffAfaAmux},
  we show resources as a function of the number of rows ($H$) stored in each image strip.
For $N=251$ and $B=8$, we also show the required number of RAM resources and the 
   total number of MUXes in Table \ref{table:Jresources_ram_mux_div}.
For comparing performance as a function of resources, 
   we present the required number of cycles
   as a function of flip-flops in Fig. \ref{fig:resourcesTime},
   and as a function of 1-bit additions in Fig. \ref{fig:JresourcesaddersTime}.

\begin{figure}[!t]
\centering
\includegraphics[width=0.5\textwidth]{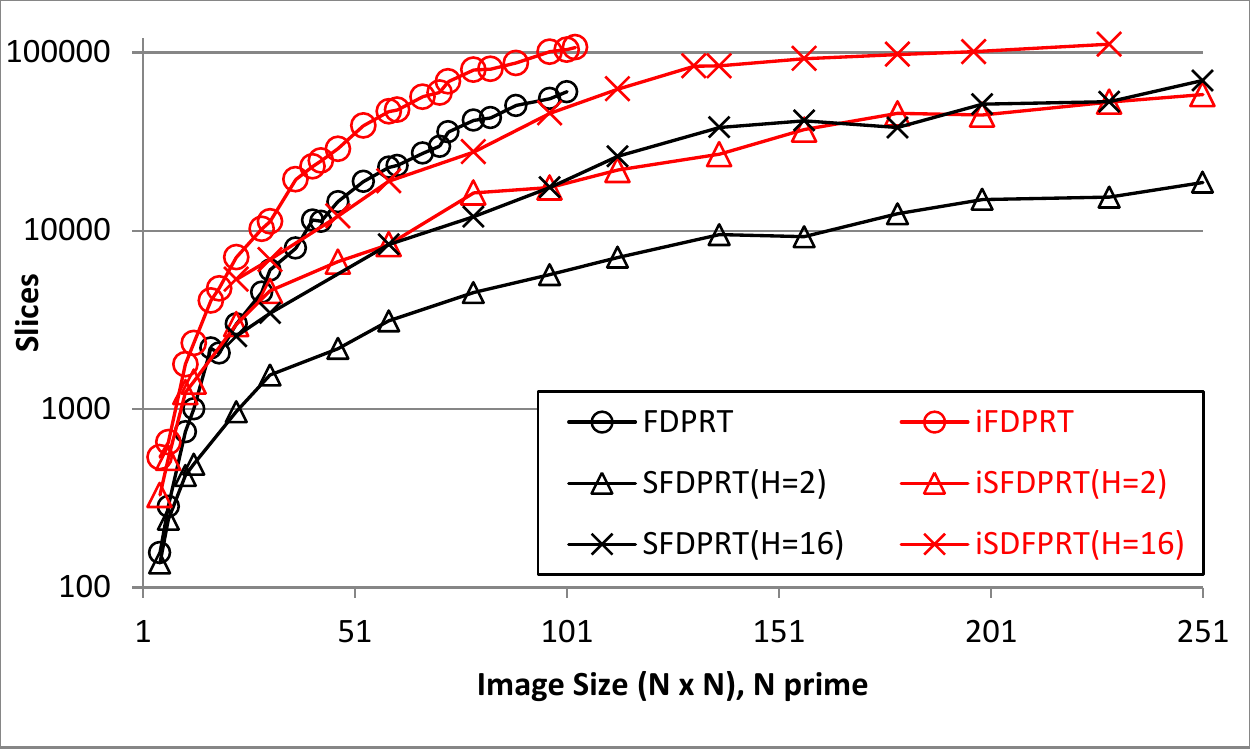} % Uses the whole column
\caption{
FPGA slices for a Virtex-6 implementation for both the forward and inverse DPRTs
   for $H=2, 16$, $N$ prime and $2\leq N\leq 251$.
}
\label{fig:FPGAresources}
\end{figure}

We present the required number of slices for a Virtex-6 implementation in Fig. \ref{fig:FPGAresources}.
As $N$ increases, we observe linear growth in the number of slices 
   as expected from our analysis in Table \ref{table:resources}.
On the other hand, for smaller values of $N$, we have quadratic growth.
The trends are due to the optimizations performed by the Xilinx synthesizer.
Overall, since Virtex-6 devices use 6-input LUTS, implementations
   that utilize all 6 inputs provide better resource optimization
   than implementations that use fewer inputs.
For the entire system, we have clock frequencies of 100 MHz for the Xilinx 6-series
    and 200 MHz for the Xilinx 7-series (Virtex-7, Artix-7, Kintex-7).

We also provide a summary of our results for the inverse DPRT.
For the fast version (iFDPRT) running time and resources, we refer back to Figs. \ref{fig:FPGArunningTime} and \ref{fig:FPGAresources}. 
For the number of input bits, we recall that $B'=B+\left\lceil \log_{2}N\right\rceil$.
Thus, overall, the iFDPRT implementations require more resources and  slightly more computational times.
Similar comments apply for the scalable, inverse DPRTs (iSFDPRT) shown in Figs. \ref{fig:FPGArunningTime} and \ref{fig:FPGAresources}.

\begin{table}
\renewcommand{\arraystretch}{1.5}
\caption{\label{table:timeforward}
Total number of clock cycles for computing the DPRT. 
In all cases, the image is of size $N \times N$, and $H=2, \ldots, N$ is the scaling factor for the SFDPRT.}
\begin{center}
\begin{tabular}{|p{0.15\textwidth}|c|}
\hline\hline
{\bf Method} & {\bf Clock cycles} \\
 \hline
 \hline
 Serial \cite{Matus1993},\cite{Chandra2005}	& $N^{3}+2N^{2}+N$  \\
 Systolic	\cite{Matus1993},\cite{Chandra2008} & $N^{2}+N+1$ \\
 \hline
{\bf Proposed Approaches:} & \\
 - SFDPRT & $\left\lceil N/H\right\rceil(N+3H+3)+N+\left\lceil \log_{2}H\right\rceil+1$  \\
 %\hfill & $+N+H+$  \\
 - SFDPRT ($H=2$) \newline \phantom{-} lowest resource use & $\left\lceil N/2\right\rceil(N+9)+N+2$ \\
 - SFDPRT ($H=N$) \newline\phantom{-} fastest running time & $5N+\left\lceil \log_{2}N\right\rceil+4$ \\
 - FDPRT	& $2N+\left\lceil \log_{2}N\right\rceil+1$ \\
\hline\hline
\end{tabular}
\end{center}
\end{table}

\begin{table}
\renewcommand{\arraystretch}{1.5}
\caption{\label{table:timeinverse}
Total number of  clock cycles for computing the iDPRT. 
Here, the image size is  $N \times N$.
We use $B$ bits per pixel, and $H=2, \ldots, N$ is the scaling factor of the iSFDPRT.
Add $N$ clock cycles in the scalable version if ${\tt MEM\_IN}$ is used.}
\begin{center}
\begin{tabular}{|p{0.14\textwidth}||c|}
 \hline\hline
Our work & Clock cycles \\
 \hline\hline
 iSFDPRT & $\left\lceil N/H\right\rceil(N+H)+2\left\lceil \log_{2}N\right\rceil$ \\
 \hfill & +$\left\lceil \log_{2}H\right\rceil+B+3$  \\
 \hline
 iSFDPRT ($H=2$) \newline lowest resource usage & $\left\lceil N/2\right\rceil(N+2)+2\left\lceil \log_{2}N\right\rceil+B+4$ \\
 \hline
 iSFDPRT ($H=N$) \newline fastest running time & $2N+3\left\lceil \log_{2}N\right\rceil+B+3$ \\
 \hline
 iFDPRT	& $2N+3\left\lceil \log_{2}N\right\rceil+B+2$ \\
\hline\hline
\end{tabular}
\end{center}
\end{table}

\begin{table*}
\renewcommand{\arraystretch}{1.5}
\caption{\label{table:resources}Resource usage for different DPRT and inverse DPRT implementations.
Here, we have  an image size of $N \times N$, $B$ bits per pixel, $n=\left\lceil \log_{2}N\right\rceil$, $h=\left\lceil \log_{2}H\right\rceil$, $K=\left\lceil N/H\right\rceil$, and $H=2, \ldots, N$. 
For the adder trees, we define ${\tt A_{ff}}$ to be number of required flip-flops,
     and ${\tt A_{FA}}$ to be the number of 1-bit additions.
For the register array, we define ${\tt A_{mux}}$ to be the number of 2-to-1 MUXes.
${\tt A_{ff}}$, ${\tt A_{FA}}$, and ${\tt A_{mux}}$ grow linearly with respect to $N$
   and can be computed using the algorithm given in 
    the appendix (Fig. \ref{alg:adderResources}).
For the inverse DPRT, we note that each divider is
    implemented using $3(B+2n)^{2}$ flip-flops, 
    $(B+2n)^{2}$ 1-bit additions, and $(B+2n)^{2}$ 2-to-1 MUXes.  
	Here, we use the term ``1-bit additions"	to refer to the number
	of equivalent 1-bit full adders.
}
\begin{center}
\resizebox{\textwidth}{!}{%
\begin{tabular}{|p{0.14\textwidth}||c|c|c|l|}
\hline\hline
\hfill & \multicolumn{4}{|c|}{Resources} \\
\cline{2-5}
\hfill & Register array & \multicolumn{2}{|c|}{Adder trees} & Others: Dividers ($B+2n$ bits) or \\
\cline{3-4}
\hfill & (in bits) & Number of flip-flops & $1$-bit additions & 	RAM(in bits), 2-to-1 MUXes\\
\hline\hline
 Serial \cite{Matus1993},\cite{Chandra2005}	& 	$N(B+n)$ & 	$3B+2n$ &	$(B+n)$ &	RAM: $N^{2} B$ \\
 \hline
 Systolic	\cite{Matus1993},\cite{Chandra2008}  & 	$N(N+1) n$ & 	$(N+1)(3B+2n)$ &	$(N+1)(B+n)$ &	RAM: $N(N+1)(B+n)$\\
 \hline
 SFDPRT & 	$N  H  B$ & 	$N {\tt A_{ff}}(H,B)$ &	$N {\tt A_{FA}}(H,B)$ &	RAM: $N^{2} B + N(N+1)(B+n)$ \\
 \hfill & \hfill & \hfill & $+N (B+n)$ & MUX: $N  H {\tt A_{mux}}(K+1,B)$\\
 \hline
 SFDPRT ($H=2$)  & 	$2N  B$ & 	$N  (B+1)$ &	$N  B$ &	RAM: $N^{2}  B + N(N+1)(B+n)$ \\
 lowest resource usage & \hfill & \hfill &  $+N  (B+n)$ & MUX: $2N {\tt A_{mux}}(\left\lceil N/2\right\rceil+1,B)$\\
  \hline
 SFDPRT ($H=N$)  & 	$N^{2}  B$ & 	$N  {\tt A_{ff}}(N,B)$ &	$N  {\tt A_{FA}}(N,B)$ &	RAM: $N^{2}  B + N(N+1)(B+n)$ \\
  fastest running time & \hfill & \hfill &  $+N (B+n)$ & MUX: $N^{2}  B$\\
 \hline
 FDPRT	 & $N^{2}  B$ & 	$N  {\tt A_{ff}}(N,B)$ &	$N  {\tt A_{FA}}(N,B)$ &	MUX: $2N^{2}  B$\\
 \hline
 iSFDPRT & 	$N  H  (B+n)$ & 	$(N+1)  {\tt A_{ff}}(H,B+n)$  &	$(N+1)  {\tt A_{FA}}(H,B+n)$  &	RAM: $N^{2}  (B+2n)$, Dividers: $N$ \\
 \hfill & \hfill & $+3N  (B+2n)$ &  $+2N  (B+2n)$ & MUX: $N  H  {\tt A_{mux}}(K+1,B+n)$\\
 \hline
 iSFDPRT ($H=2$)  & 	$2N  (B+n)$ & 	$(N+1)  (B+n+1)$ &	$(N+1)  (B+n)$ &	RAM: $N^{2}  (B+2n)$, Dividers: $N$\\
 lowest resource usage & \hfill & $+3N  (B+2n)$ &  $+2N  (B+2n)$ &  MUX: $2N {\tt A_{mux}}(\left\lceil N/2\right\rceil+1,B+n)$\\
  \hline
 iSFDPRT ($H=N$)  & 	$N^{2}  (B+n)$ & 	$(N+1)  {\tt A_{ff}}(N,B+n)$ &	$(N+1)  {\tt A_{FA}}(N,B+n)$ &	RAM: $N^{2}  (B+2n)$, Dividers: $N$\\
  fastest running time & \hfill & $+3N  (B+2n)$ & $+2N  (B+2n)$ & MUX: $N^{2}  (B+n)$\\
 \hline
 iFDPRT	 & $N^{2}  (B+n)$ & 	$(N+1)  {\tt A_{ff}}(N,B+n)$ &	$(N+1)  {\tt A_{FA}}(N,B+n)$ &	Dividers: $N$ \\
  \hfill & \hfill & $+N  (B+2n)$ & $+N  (B+2n)$ & MUX: $N^{2}  (B+n)$\\
 \hline\hline
\end{tabular}}
\end{center}
\end{table*}

\begin{table}
	\renewcommand{\arraystretch}{1.5}
	\caption{\label{table:Jresources_ram_mux_div}
		Total number of resources for RAM (in 1-bit cells) and MUXes (2-to-1 muxes). 
		The resources are shown for $N = 251$. 
		Except for the MUXes for the SFDPRT, the values refer to any $H$.
		The number of MUXes for the SFDPRT refer to values of $H$ that
		lie on the Pareto front$ ^*$.
	}
	\begin{center}
		\begin{tabular}{|p{0.15\textwidth}|c|c|}
			\hline\hline
			{\bf Method} & {\bf RAM} & {\bf MUXes} \\
			\hline
			\hline
			Serial \cite{Matus1993},\cite{Chandra2005}	& $504,008$ & Unknown  \\
			Systolic	\cite{Matus1993},\cite{Chandra2008} & $1,012,032$ & Unknown \\
			\hline
			SFDPRT & $1,516,040$ & $506,016^*$  \\
			FDPRT	& $0$ & $1,008,016$ \\
			\hline\hline
		\end{tabular}
	\end{center}
\end{table}

\begin{comment}
\begin{table}
\renewcommand{\arraystretch}{1.5}
\caption{\label{table:FPGAresources}Resource usage for different FPGAS implementations.}
\begin{center}
\begin{tabular}{|c|c|c|c|}
\hline\hline
Design & Size ($N$) & Slices & Platform \\
\hline\hline
 Serial \cite{Matus1993},\cite{Chandra2005} & 5,7,11,17 & 248,350,548,972 & Virtex II \\
 \hline
 Systolic	\cite{Matus1993},\cite{Chandra2008}	& 7 & 245 & Virtex II \\
 \hline
 FDPRT	& 5,7,11,17 & 434,834,2221,5304 & Virtex IV \\
 \hline
 FDPRT	& 5,7,11,17 & 156,285,745,2194 & Virtex VI \\
 \hline
 SFDPRT(H=2)	& 5,7,11,17 & 137,240,424,826 & Virtex VI \\
 \hline
 SFDPRT(H=4)	& 5,7,11,17 & 212,272,455,925 & Virtex VI \\
 \hline
 SFDPRT(H=9)	& 17 & 1252 & Virtex VI \\
 \hline
 SFDPRT(H=17)	& 17 & XXX & Virtex VI \\
\hline\hline
\end{tabular}
\end{center}
\end{table}
\end{comment}

\subsection{Discussion}\label{sec:discussion}
Overall, the proposed approach results in the fastest running times. 
Even in the slowest cases, our running times are significantly better than 
   any previous implementation.
Scalable DPRT computation has also been demonstrated
  where the required number of cycles can be reduced
  when more resources are available.
Significantly faster DPRT computation is possible
   for fixed size transforms when the architecture
    can be implemented using available resources.
Furthermore, these results have been extended for the inverse DPRT.   
However, in some cases, the better running times come at a cost of increased resources.
Thus, we also need to discuss how our running times depend on the number of required resources.

For an $N\times N$ image ($N$ prime), the proposed
  approaches can compute the DPRT in significantly
  less time than $N^2$ cycles.    
The fastest architectures (FDPRT and iFDPRT) compute the 
  forward DPRT and inverse DPRT in
  just $2N+\left\lceil \log_{2}N\right\rceil+1$ and  
  $2N+3\left\lceil \log_{2}N\right\rceil+B+2$ cycles respectively 
  (where $B$ is the number of bits used to represent each input pixel).
When resources are available, the scalable approach can 
  also compute the DPRT in a number of cycles that 
  is linear in $N$.
In the fastest case, the scalable DPRT   
  requires $2N+\left\lceil \log_{2}N\right\rceil+1$ clock cycles.
However, when very limited resources are available, the 
  number of required clock cycles
  increases to $\left\lceil N/2\right\rceil (N+9)+N+2$
  for the case where we only have two image rows per strip $H=2$.

Based on Fig. \ref{fig:resourcesTime},
  we compare the number of cycles as a function of the required number of flip-flops.  
From the Figure, we note that systolic implementation requires 
  $516,096$ flip flops to compute the DPRT for a $251 \times 251$ image in $63,253$ clock cycles (square dot in Fig. \ref{fig:resourcesTime}).
In comparison, with $25\%$ less resources for $H=84$, we have that the scalable DPRT 
  is computed  36 times faster than the systolic implementation.
On the other hand, for the serial implementation, we note that the proposed scalable 
  DPRT approaches are much faster but require more resources.
The fast DPRT implementation requires only 511 cycles that is vastly superior to any other approach.

Based on Fig. \ref{fig:JresourcesaddersTime}, 
  we also compare the number of cycles as a function of the number of 1-bit additions.
As expected, the serial implementation requires a single 16-bit adder.
However, the serial implementation is very slow compared to all other implementations.
The systolic implementation requires only 4,032 1-bit additions that 
  is close to the two-row per strip ($H=2$) implementation of the scalable DPRT.
However, in all cases, the systolic implementation is significantly slower
  than all of the proposed implementations.
Essentially, 
   the scalable approach improves its 
   performance while requiring more 1-bit additions
   for larger values of $H$.

As detailed in section \ref{sec:Pareto}, we are only interested
  in Pareto-optimal implementations.
Here, the Pareto-optimal cases represent scalable implementations
  that always improve performance by using more resources.
The collection of all of the Pareto-optimal implementations
  form the Pareto-front and are shown in Fig. \ref{fig:resourcesTime} and Fig. \ref{fig:JresourcesaddersTime}.

The proposed system can also be expanded for use in FPGA co-processor systems
where the FPGA card communicates with the CPU using a PCI express interface.
Clearly, the advantage of using the proposed architecture increases with $N$ since
the image transfer overhead will not be significant for larger $N$.
To understand the limits, assume a PCI express 3.x bandwidth of about 
16 GB/s and a general-purpose microprocessor that achieves the maximum performance of 
10 Giga-flops using 4 cores at 2.5 GHz.
In terms of CPU memory accesses, we assume a 32GB/second bandwidth for DDR3 memory.
Furthermore, suppose that we are interested in computing the DPRT of a $251\times 251$ image.    
In this case, image I/O requires about 3.67 micro-seconds per image transfer
from the DDR3 to the FPGA card.
The DPRT requires      
$N^2*(N+1)\approx 15.88$ mega floating point operations 
for the additions.
The additions can be performed in 1.479 milli-seconds (1479 micro-seconds) on the CPU.    
In addition, the CPU implementation will need 
$2 N^2$ DDR3 memory accesses for implementing the transposition and
retrieving the matrix in shifted form.
However, assuming that these memory accesses are implemented effectively using
DDR3 memory, that only requires 3.67 micro-seconds.
Hence, the CPU computation will be dominated by the additions.
On the other hand, DPRT computation on an FPGA operating at just 100 MHz for older devices 
(at half the 200 MHz of more modern FPGA devices),
will only require 2*251+9 cycles in about 5.11 micro-seconds.
Thus, the speedup factor is above $(3.67+1479)/(3.67+5.11) \approx 169$.

We also consider comparisons of the DPRT to
previous implementations of the Hough transform.
Here, we note that the Hough transform can be used
to detect lines by adding up edge pixels 
along different directions.
In this application, the Hough transform
is computed using the discrete Radon transform (DRT).
As noted earlier, prior to adding up values along different directions, 
typical implementations of the DRT require interpolation.
Furthermore, the DRT is defined over the original image
as opposed to its periodic extension as required by the DPRT.
Despite these limitations of our comparisons, we note that 
the proposed DPRT implementations are substantially faster
than the fastest Hough transform implementations.
For example, the authors of \cite{chen2012resource}
report on an FPGA implementation, operating at 200 MHz, that requires 2.07-3.16 ms
for detecting lines over 180 orientations in 
$512\times 512$ images.     
Similar comments apply to related to continuous-space extension of the 
Radon transform (e.g., generalized, hyperbolic, parabolic, etc.).

\section{Conclusions and Future Work}\label{sec:conclusions}
The manuscript summarized the development of 
   fast and scalable methods for computing the DPRT and its inverse.
Overall, the proposed methods provide much faster computation of the DPRT.
Furthermore, the scalable DPRT methods provide fast execution times that can be 
   implemented within available resources. 
In addition, we present fast DPRT methods that provided the fastest execution times 
   among all possible approaches.
For an $N\times N$ image, the fastest DPRT implementations
   require a number of cycles that grows linearly with $N$.   
Furthermore, in terms of resources, the proposed 
  architectures only require fixed point additions and shift registers.

Currently, we are working on the application of the DPRT for computing
   fast convolutions.
As with the current manuscript, our focus will be to extend the current
   DPRT architecture so as to support the multiplication
   of the DPRT of the input image with the impulse response of
   a larger filter, and then take the inverse.
Beyond the FPGA implementation, we are also developing GPU implementations.
Future work will also focus on the development of 
   fast methods for computing 2-D Discrete Fourier Transforms (DFTs).

\section{Acknowledgments} 
This material is based upon work supported by
the National Science Foundation under NSF AWD CNS-1422031.

% or
\appendix  % for no appendix heading


\begin{figure}[h!]
\begin{algorithmic}[1]
\Procedure {{\tt Tree\_Resources}}{$X,B$}
\State $h=\left\lceil \log_{2}X\right\rceil$
\State ${\tt A_{ff}} = {\tt A_{FA}} = {\tt A_{mux}} = 0$
\State $a=X$
\For {$z$ = $1$ to $h$}
	\State $r = \left\langle a\right\rangle _{2}$
	\State $a = \left\lfloor a/2\right\rfloor$
	\State ${\tt A_{FA}} = {\tt A_{FA}} + a \cdot (B+z-1)$
	\State ${\tt A_{mux}} = {\tt A_{mux}} + a \cdot B$
	\State $a = a + r$
	\State ${\tt A_{ff}} = {\tt A_{ff}} + a \cdot (B+z)$
\EndFor
\State \textbf{return} ${\tt A_{FA}}, {\tt A_{ff}}, {\tt A_{mux}}$
\EndProcedure
\end{algorithmic}
\caption{\label{alg:adderResources}
Required tree resources as a function of the number of strip rows or number of blocks ($X$),
    and the number of bits per pixel ($B$).
Refer to Table \ref{table:resources} for definitions of   ${\tt A_{ff}}, {\tt A_{FA}}, {\tt A_{mux}}$.
For ${\tt A_{ff}}$, the resources do not include the input registers,
    but do include the output registers since they are
    implemented in ${\tt SFDPRT\_core}$ and ${\tt iSFDPRT\_core}$.
}
\end{figure}

\ifCLASSOPTIONcaptionsoff
  \newpage
\fi



%\bibliographystyle{IEEEtran}

\bibliographystyle{IEEEtran} % Inserted by CCD, also in the working dir add: IEEEtran.bst and Conv_Radon.bib

% ???????????????? Marios changed the bibliography
\bibliography{Conv_Radon} % Inserted by CCD


%\begin{thebibliography}{1}
%\end{thebibliography}


\begin{IEEEbiography}[{\includegraphics[width=1in,height=1.25in,clip,keepaspectratio]{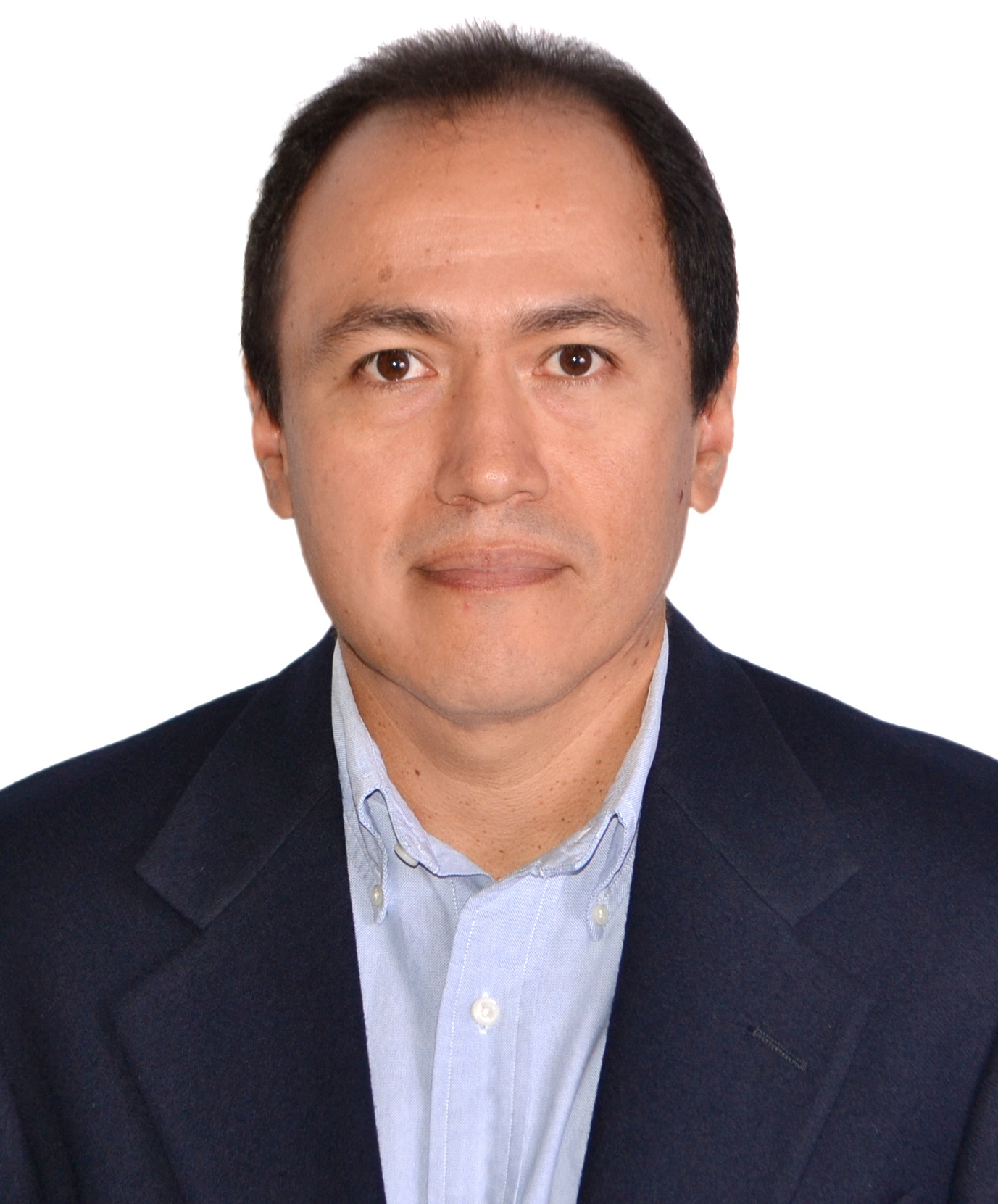}}]{Cesar Carranza}
Cesar Carranza is a Ph.D. candidate in Computer Engineering and received his M.Sc. degree in
Computer Engineering in 2012 from the University of New Mexico at Albuquerque. He also
holds a M.Sc in Computer Science from Centro de Investigaci\'on Cient\'ifica y de Educaci\'on Superior de Ensenada in 2010 and a B.Sc. in Electrical Engineering from Pontificia Universidad Cat\'olica del Per\'u in 1994.
He is currently an Assistant Professor at Pontificia Universidad Cat\'olica del Per\'u. His current research interests include parallel algorithms for image processing, high performance hardware integration and parallel computing.
\end{IEEEbiography}

\begin{IEEEbiography}[{\includegraphics[width=1in,height=1.25in,clip,keepaspectratio]{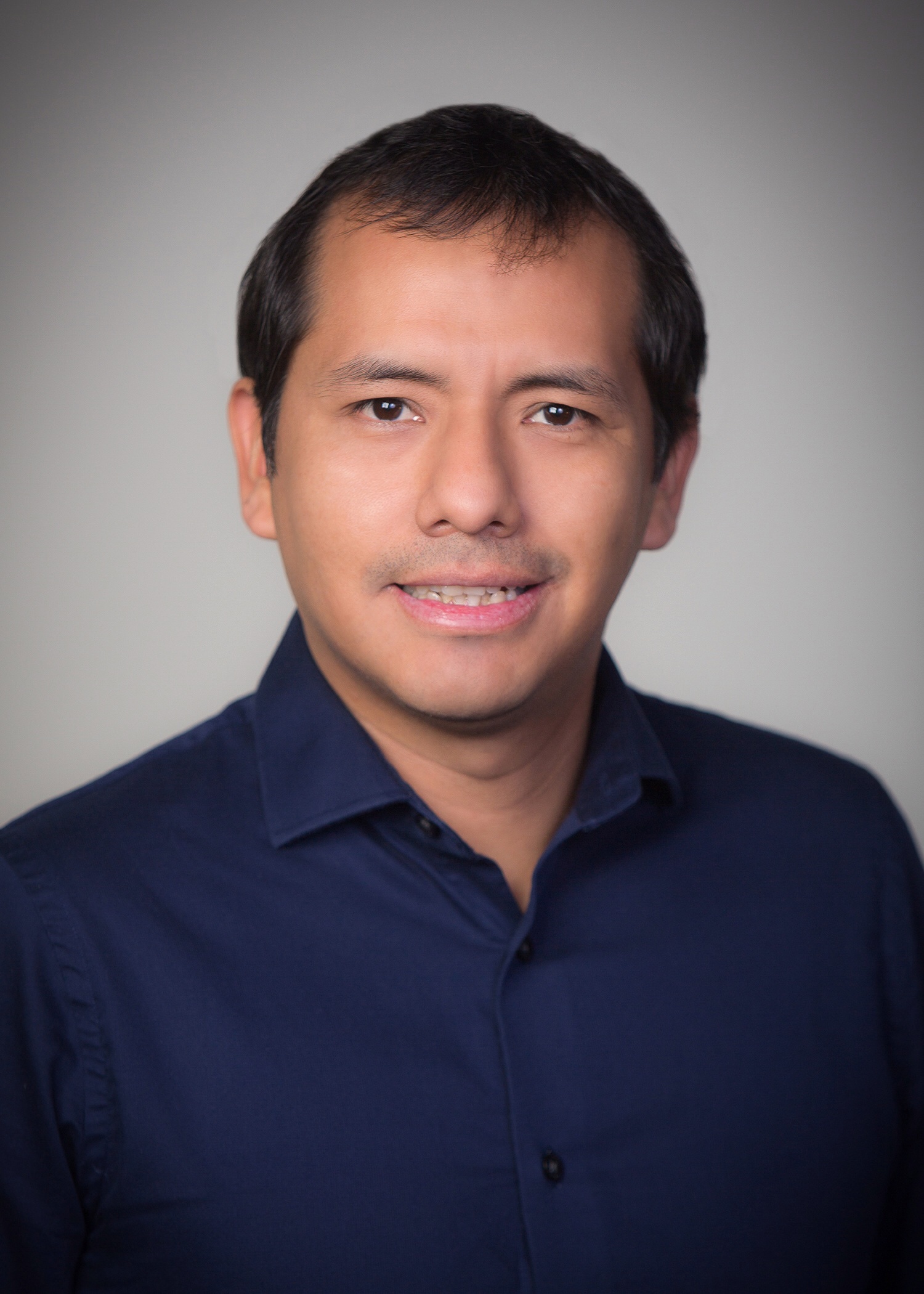}}]{Daniel Llamocca}
Daniel Llamocca received his Ph.D. degree in Computer Engineering and his M.Sc. degree in Electrical Engineering from the University of New Mexico at Albuquerque in 2012 and 2008 respectively. He also holds a B.Sc. in Electrical Engineering from Pontificia Universidad Cat\'olica del Per\'u in 2002.

He is currently an Assistant Professor at Oakland University. His research deals with run-time automatic adaptation of hardware resources to time-varying constraints with the purpose of delivering the best hardware solution at any time. His current research interests include: i) reconfigurable computer architectures for signal, image, and video processing, ii) high-performance architectures for computer arithmetic, communication, and embedded interfaces, iii) embedded system design, and iv) Run-time Partial Reconfiguration techniques on FPGAs.
\end{IEEEbiography}

\begin{IEEEbiography}[{\includegraphics[width=1in,height=1.25in,clip,keepaspectratio]{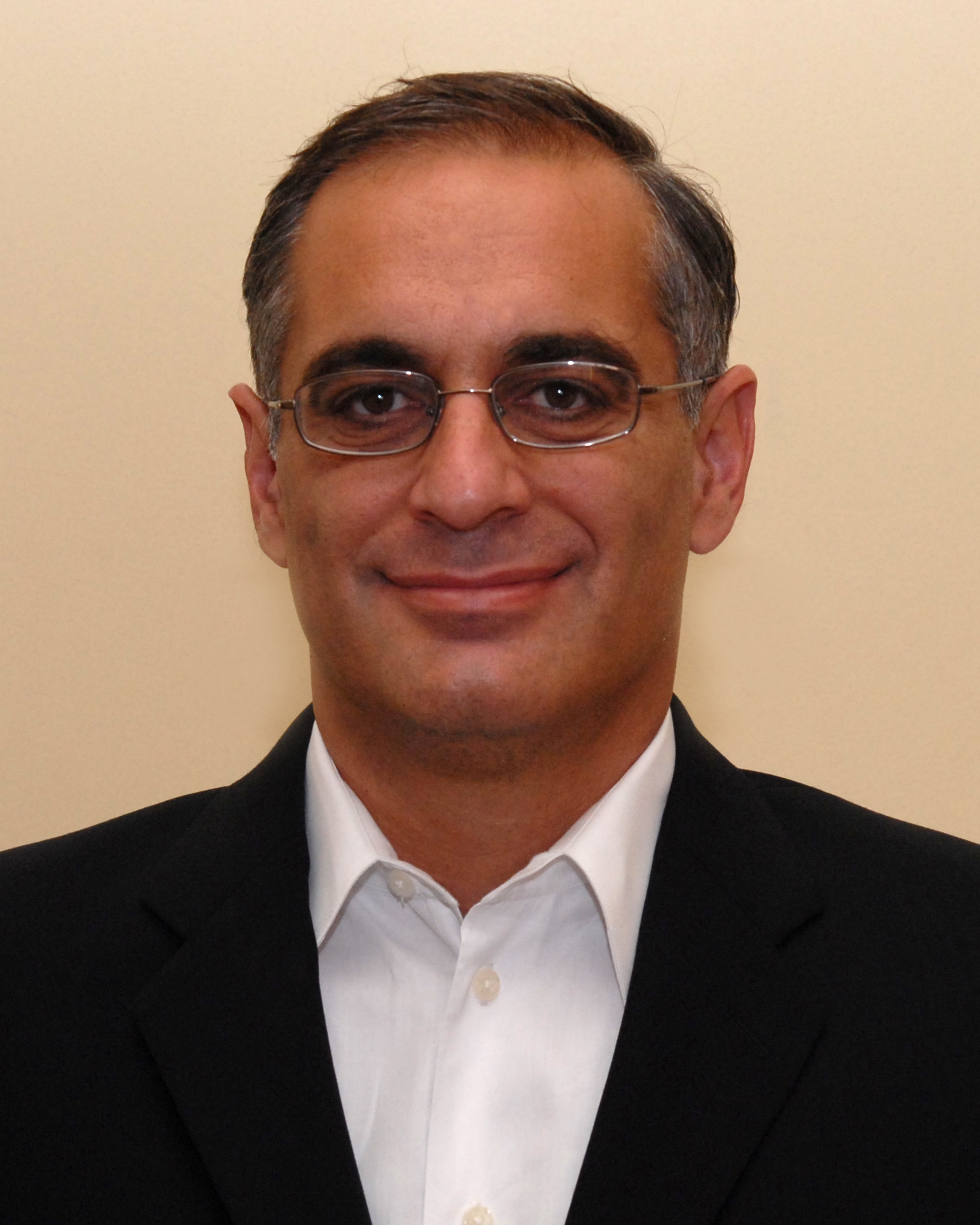}}]{Marios Pattichis}
Marios Pattichis (M'99, SM'06) received the B.Sc.(High Hons. and Special Hons.) degree in computer sciences and the B.A. (High Hons.) degree in mathematics, both in 1991, the M.S. degree in electrical engineering in 1993, and the Ph.D. degree in computer engineering in 1998, all from the University of Texas, Austin. He is currently a Professor with the Department of Electrical and Computer Engineering, University of New Mexico (UNM), Albuquerque. His current research interests include digital image, video processing, communications, dynamically reconfigurable computer architectures, and biomedical and space image-processing applications.

Dr. Pattichis is currently a senior associate editor of the {\it IEEE Signal Processing Letters}.
He has served as an associate editor for the {\it IEEE Transactions on Image Processing},
{\it IEEE Transactions on Industrial Informatics}, and has also served as a guest associate editor for the {\it IEEE Transactions on Information Technology in Biomedicine}. He was the general chair of the {\it 2008 IEEE Southwest Symposium on Image Analysis and Interpretation}. He was a recipient of the 2004 Electrical and Computer Engineering Distinguished Teaching Award at UNM. For his development of the digital logic design labs at UNM he was recognized by the Xilinx Corporation in 2003 and by the UNM School of Engineering's Harrison faculty excellent award in 2006. He was a founding Co-PI of COSMIAC at UNM. At UNM, he is currently the director of the image and video Processing and Communications Lab (ivPCL). % , \url{ivpcl.org}).
\end{IEEEbiography}


%\vfill


\end{document}